\documentclass[namedreferences]{solarphysics}
\usepackage[optionalrh,solaromanenum,linksfromyear,natbib]{spr-sola-addons} 
\usepackage{graphicx}                    
\usepackage{amssymb}                    
\usepackage{color}                       
\usepackage{url}                         
\usepackage{hyperref}
\usepackage[normalem]{ulem}	
\hypersetup{colorlinks, citecolor=blue, filecolor=black, linkcolor=black, urlcolor=black}

\usepackage{bm}
\makeatletter

\newcommand{\Rmnum}[1]{\expandafter\@slowromancap\romannumeral #1@}
\makeatother

\newcommand{\corr}[1]{{{#1}}}

\begin{document}

\begin{article}

\begin{opening}

\title{\textsf{CorPITA}: An Automated Algorithm for the Identification and Analysis of Coronal ``EIT Waves''}

%
\author{D.M.~\surname{Long}$^{1}$\sep
        D.S.~\surname{Bloomfield}$^{2}$\sep
        P.T.~\surname{Gallagher}$^{2}$\sep
        D.~\surname{P\'{e}rez--Su\'{a}rez}$^{3}$
       }

%
\runningauthor{D.M.~Long \emph{et al.}}
\runningtitle{Automated ``EIT wave'' Analysis Using CorPITA}

%
  \institute{$^{1}$ Mullard Space Science Laboratory, University College London, Holmbury St. Mary, Dorking, Surrey, RH5 6NT, UK\\
                     email: \url{david.long@ucl.ac.uk} \\ 
             $^{2}$ School of Physics, Trinity College Dublin, College Green, Dublin 2, Ireland \\
             $^{3}$ South African National Space Agency (SANSA) Space Science, 7200 Hermanus, South Africa
             }

\begin{abstract}
The continuous stream of data available from the \corr{\emph{Atmospheric Imaging Assembly}} (AIA) telescopes onboard the \corr{\emph{Solar Dynamics Observatory}} (SDO) spacecraft has allowed a deeper understanding of the Sun. However, the sheer volume of data has necessitated the development of automated techniques to identify and analyse various phenomena. In this \corr{article}, we describe the Coronal Pulse Identification and Tracking Algorithm (\textsf{CorPITA}) for the identification and analysis of coronal ``EIT waves''. \textsf{CorPITA} uses an intensity-profile technique to identify the propagating pulse, tracking it throughout its evolution before returning estimates of its kinematics. The algorithm is applied here to a data-set from February~2011, allowing its capabilities to be examined and critiqued. This algorithm forms part of the SDO Feature Finding Team initiative and will be implemented as part of the Heliophysics Event Knowledgebase (HEK). This is the first fully automated algorithm to identify and track the propagating ``EIT wave'' rather than any associated phenomena and will allow a deeper understanding of this controversial phenomenon.
\end{abstract}

%

\end{opening}

%
\section{Introduction} \label{sect:intro} 

Solar eruptions are the most energetic events in our solar system, releasing large bursts of radiation as solar flares and ejecting plasma into the heliosphere as coronal mass ejections (CMEs). On the Sun, these eruptions are often associated with large-scale disturbances that propagate across the solar atmosphere at typical speeds of $\approx$200\,--\,400~km~s$^{-1}$ \citep{Thompson:2009il} although more recently velocities of up to $\sim$1500~km~s$^{-1}$ have also been measured \citep[\emph{e.g.}][]{Olmedo:2012ff,Shen:2013tw}. Initially observed by \citet{Moses:1997qa}, \citet{Dere:1997fk} and \citet{Thompson:1999zt} using the \corr{\emph{Extreme ultraviolet Imaging Telescope}} \citep[EIT:][]{Delaboudiniere:1995ve}, these disturbances (commonly called ``EIT waves'') have been studied in detail for more than $\approx$15~years. However, they remain a source of debate with conflicting observations of their properties leading to a myriad of theories proposed to explain the phenomenon.

Initial studies of the ``EIT wave'' feature interpreted it as a fast-mode magnetoacoustic wave using the theory originally proposed by \citet{Uchida:1968gb,Uchida:1970jl}, with the disturbance initiated by the same eruption producing the associated solar flare and CME. This is consistent with observations of refraction \citep{Veronig:2006cq} and reflection \citep{Gopalswamy:2009kl} at coronal-hole boundaries and observed pulse properties such as pulse dispersion and dissipation \citep[\emph{e.g.}][]{Long:2011fv,Muhr:2011pi}, although some authors have suggested alternate wave interpretations such as solitons \citep{Wills-Davey:2007mw} or slow-mode MHD waves \citep{Podladchikova:2010fu}. However, observations of stationary bright fronts at coronal-hole boundaries and low pulse speeds have lead to the proposal of ``pseudo-wave'' theories. These interpretations see the disturbance not as a true wave, but as a bright feature produced by Joule heating at the boundary between the erupting CME and the background coronal magnetic field as the CME propagates into the heliosphere \citep{Delannee:2000bh,Delannee:2008qf}. A third alternative, originally proposed by \citet{Chen:2002zr}, combines aspects of both the wave and pseudo-wave theories to interpret ``EIT waves'' as a hybrid of both. In this case, the erupting CME drives a fast-mode wave \corr{that} then propagates freely, while magnetic reconnection driven by the restructuring magnetic field as the CME erupts is seen as a second, slower propagating feature. This theory was built upon in a subsequent \corr{article} by \citet{Chen:2005ys} and has been the focus of further simulations performed by \citet{Cohen:2009ly} and \citet{Downs:2011nx,Downs:2012cr}. There has also been some observational evidence for a second propagating front, particularly \corr{in work} by \citet{Zhukov:2004if}, \citet{Chen:2011vn}\corr{,} and \citet{Harra:2011hc}. A detailed discussion of the different proposed theories and the evidence for and against them may be found in the recent reviews by \citet{Wills-Davey:2009qc}, \citet{Gallagher:2011oq}\corr{,} and \citet{Zhukov:2011ud}.

The rich variety of theories proposed to explain the ``EIT wave'' phenomenon can be explained by the methods typically used to study them. As relatively rare events\corr{,} ``EIT waves'' are often studied in isolation, with single-event studies used to infer general phenomenological properties. Despite more than 15~years of analysis, this approach remains the primary technique for investigating them and can explain the level of uncertainty that still surrounds their true physical nature. However, several authors have proposed larger statistical surveys of the phenomenon, with the work of \citet{Thompson:2009il} in particular the benchmark for statistical analysis of ``EIT waves''. \citet{Thompson:2009il} manually identified 176 events observed by EIT onboard the \emph{SOlar and Heliospheric Observatory} \citep[SOHO:][]{Domingo:1995dq} spacecraft between 24~March~1997 and 24~June~1998, finding speeds ranging from $\approx$\,50\,--\,700~km~s$^{-1}$, with values most typically being $\approx$\,200\,--\,400~km~s$^{-1}$. This catalogue has since been utilised by multiple authors, with \citet{Biesecker:2002uq} using it to show an unambiguous correlation between ``EIT waves'' and CMEs while more recent work by \citet{Warmuth:2011kh} used it to show evidence for three distinct kinematic classes of ``EIT waves''. 

A recent \corr{article} by \citet{Nitta:2013kc} can be seen as the spiritual successor to the work of \citet{Thompson:2009il}, producing a catalogue of ``EIT waves'' observed by the \corr{\emph{Atmospheric Imaging Assembly}} \citep[AIA:][]{Lemen:2012bs} onboard the \emph{Solar Dynamics Observatory} \citep[SDO:][]{Pesnell:2012lh} spacecraft. In this case, 171 disturbances were manually identified between April~2010 and January~2013, with the observations used to examine the relationship between ``EIT waves'', solar flares and CMEs although no relationship was found between the wave speed and flare intensity or CME magnitude.

Despite their breadth and impact, the catalogues compiled by both \citet{Thompson:2009il} and more recently \citet{Nitta:2013kc} consist of manually identified events, with the consequence that identification is entirely user-dependent. This was noted by \citet{Thompson:2009il}, who assigned one of six quality ratings to each event as ``an indicator of the observability of the wave in the data''. However, the rating is entirely subjective and dependent on consistent application by the authors. This approach means that the same parameters may not necessarily be applied consistently to events identified by both catalogues, despite best efforts.

Several authors have proposed automated approaches to the identification and analysis of ``EIT waves'' to try \corr{to} overcome these issues. The Novel EIT wave Machine Observing (NEMO) catalogue was developed by \citet{Podladchikova:2005ye} to identify and analyse the coronal dimmings associated with ``EIT waves'' using data from SOHO/EIT (\url{sidc.oma.be/nemo/}). This technique was successfully implemented and operated at the Solar Influences Data Analysis Centre at the Royal Observatory of Belgium from 1997 to 2010 and primarily identified coronal-dimming regions, although a strong correlation was noted between the dimming regions and the propagating ``EIT wave'' feature. An alternative technique using Huygens tracking was proposed by \citet{Wills-Davey:2006ss}. This approach employs percentage base-difference \citep[PBD; cf.][]{Wills-Davey:1999fc} images, with the pulse identified by finding the line of peak intensities corresponding to the peak of the Gaussian cross-section of the pulse. Once the disturbance has been identified in each image, its path of propagation is found using a reverse-engineered Huygens tracking approach. While highlighting several of the issues associated with the manual identification of ``EIT waves'' and attempting to rectify them, this technique has not been implemented on a larger scale to the best of the authors' knowledge.

As noted by \citet{Aschwanden:2010fk}, the sheer volume of data available from SDO/AIA has underlined the need for automated algorithms to identify and characterise events of interest to the wider solar community. The SDO \corr{\emph{Feature Finding Team}} \citep[FFT:][]{Martens:2012kl} has produced and implemented a number of automated techniques designed to identify and analyse features ranging from active regions and coronal holes to flares, coronal bright points, waves, CMEs, and filaments. In this \corr{article}, we discuss the Coronal Pulse Identification and Tracking Algorithm (\textsf{CorPITA}), the ``EIT wave'' detection module for the FFT. This algorithm is designed to identify, track, and analyse ``EIT waves'' using the continuous data stream available from SDO/AIA. The algorithm is outlined in Section~\ref{sect:alg}, before being applied to a sample data set in Section~\ref{sect:app}. The performance and caveats of the algorithm are then discussed and conclusions drawn in Section~\ref{sect:conc}.

\section{Algorithm Overview} \label{sect:alg}

\begin{figure*}[!t]
\begin{center}
\includegraphics[keepaspectratio,width=1\textwidth,trim=100 70 100 15,clip=]{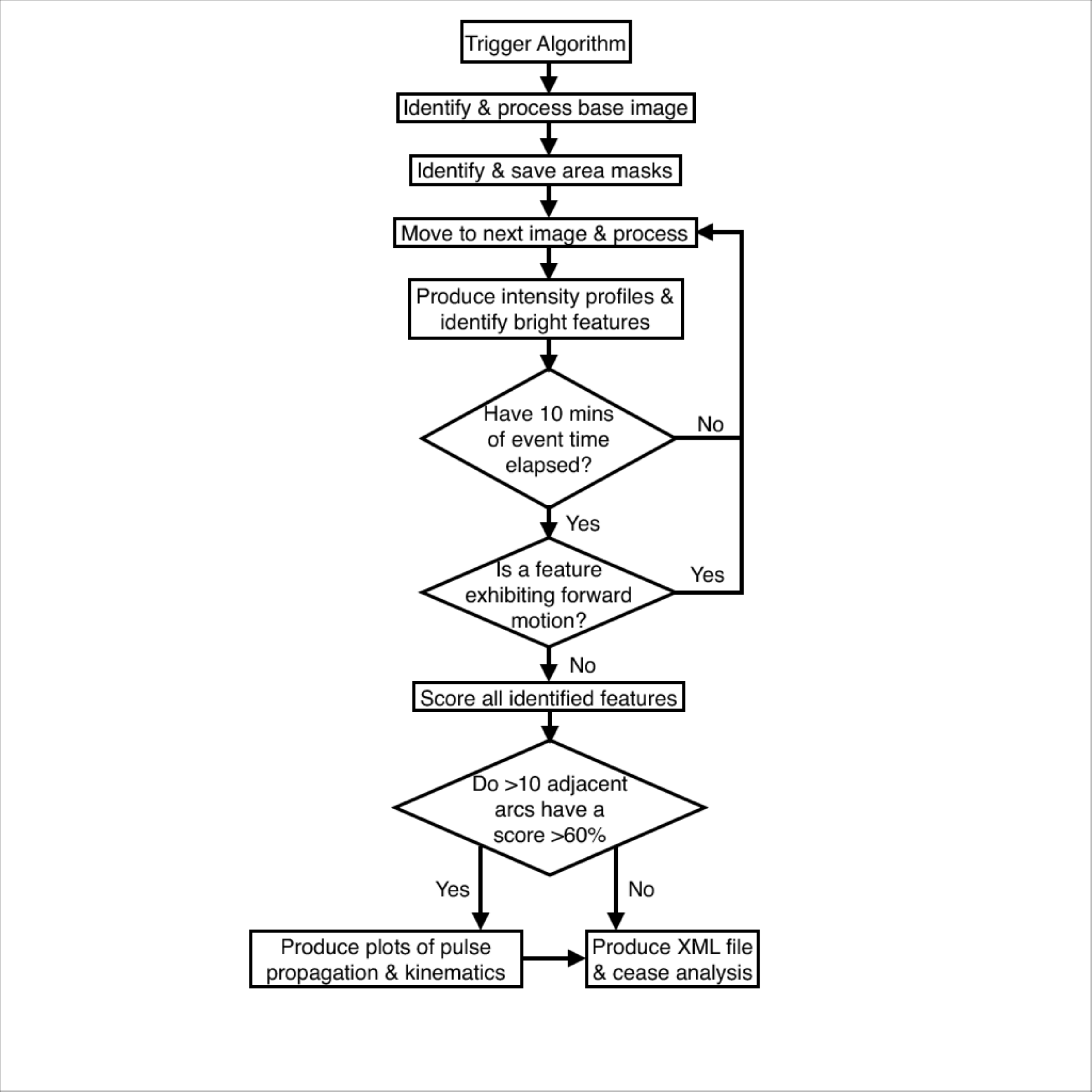}
\end{center}
\caption{A flowchart describing the basic operation of the \textsf{CorPITA} algorithm. A more detailed outline may be found in the text.}
\label{fig:pseudo_code}
\end{figure*}

The transient nature of coronal ``EIT waves'' means that \textsf{CorPITA} is designed to operate as a triggered algorithm\corr{, using the 211~\AA\ passband to identify the pulse. Although ``EIT waves'' are easily identifiable in both the 211 and 193~\AA\ passbands, the background emission from the solar corona is much lower in the hotter 211~\AA\ passband, making automated identification and tracking of the pulse much easier.} A source point is defined, with intensity profiles created along a series of arcs radiating from that point, which are then used to identify the propagating pulse. The pulse is tracked both spatially across the arcs and temporally along the arcs as it propagates across the Sun, with the kinematics of the pulse calculated once it can no longer be identified. There are therefore several distinct parts to the analysis, including the initial preparation, the image processing, and finally the detection and analysis of the event. These are outlined in detail in the sections below \corr{and are shown graphically in the flowchart depicted in Figure~\ref{fig:pseudo_code}}.

\subsection{Event Preparation}\label{subsect:prep}

\textsf{CorPITA} was designed to operate as a triggered rather than synoptic algorithm as a result of the random, transient nature of coronal ``EIT wave'' pulses. It enables calculation of the pulse kinematics, which require measuring the temporal variation of the distance of the pulse from some point. To do this, the algorithm is triggered when a solar flare erupts, allowing the location of the flare to be taken as the source from which the propagation distance of the pulse may be measured. Note that this does not imply a causal relationship between the flare and the propagating pulse, instead ensuring that events are treated in a consistent manner and allowing the uncertainty with regard to the actual source of the pulse to be determined from the initial identification of the pulse.

\begin{figure*}[!t]
\begin{center}
\includegraphics[keepaspectratio,width=1\textwidth,trim=10 15 10 15]{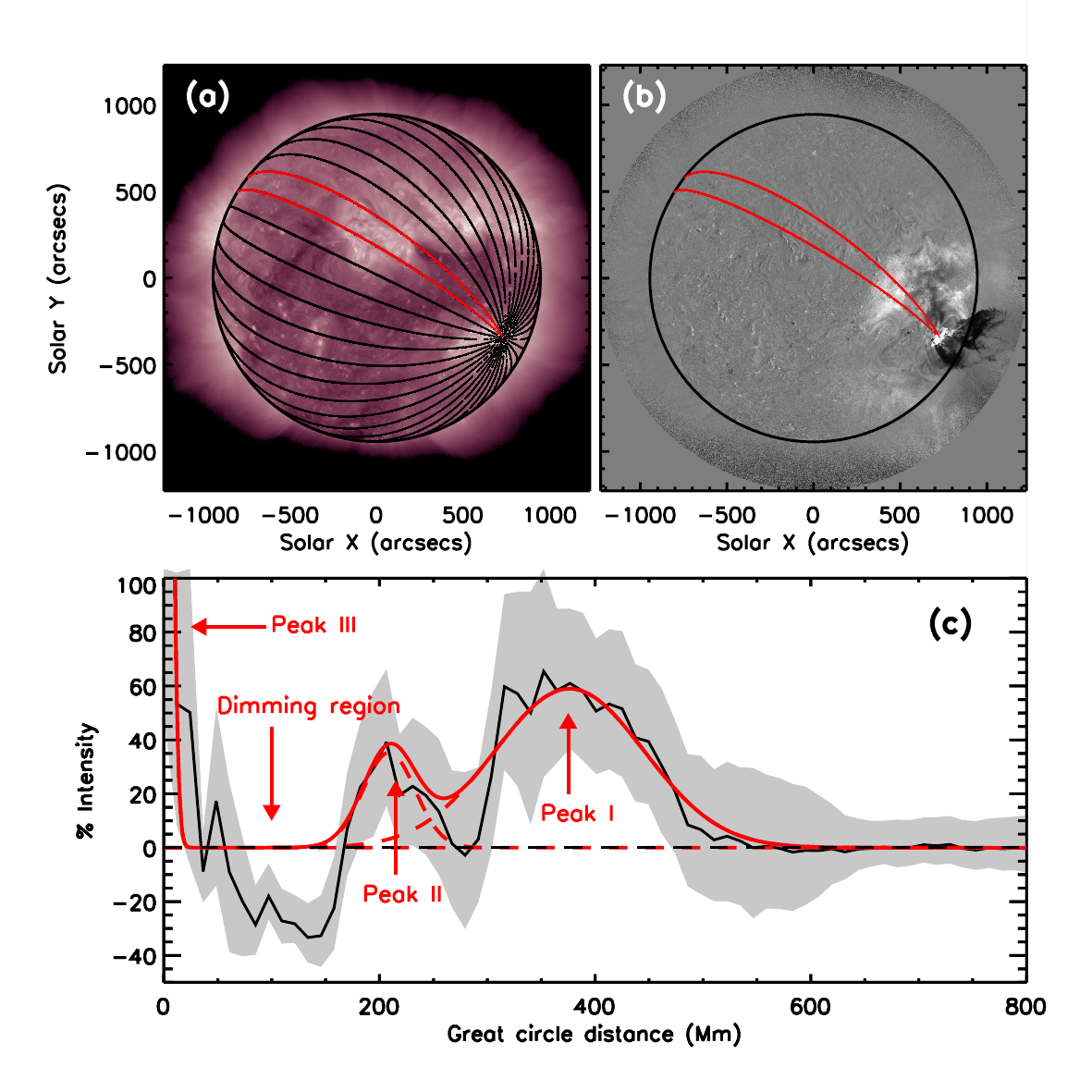}
\end{center}
\caption{Panel a shows the base image from 7~June~2011 with an example of the \corr{area masks} used by \textsf{CorPITA} to produce intensity profiles. Each \corr{area} shown here has a width of 10$^{\circ}$. Panel b shows the resulting percentage base-difference image from 06:35:03~UT on 7~June~2011, with the on-disk ``EIT wave'' seen as a bright feature. Panel c shows the resulting \corr{mean} intensity profile \corr{(in black)} produced \corr{from} the \corr{area bounded} in red in panels a \corr{and} b (305$^{\circ}$ clockwise from Solar North). The three bright features identified by the algorithm \corr{(displayed in red)} are indicated by arrows, with the shaded region indicating the \corr{uncertainty of} the intensity profile \corr{determined by the standard deviation}.}
\label{fig:wave_prof}
\end{figure*}

Once triggered, a pre-event image is identified as the image 120~seconds prior to the flare start time, with the time period chosen to minimise contamination due to the flare. This pre-event image is then taken as a base image to be used in conjunction with all subsequent images to produce percentage base-difference images that will be used to identify the pulse (see Figure~\ref{fig:wave_prof}b for an example). The percentage base-difference images are produced using,
\begin{equation}
I(t) = \left(\frac{I_t - I_0}{I_0}\right)\times 100,
\end{equation}
where $I_t$ is the image intensity at time $t$ and $I_0$ is the pre-event image intensity. PBD images are used as they allow changes in the pulse morphology and intensity to be determined relative to a consistent background coronal intensity. This is important for measuring the \corr{true} variation in pulse width and peak intensity with propagation as the \corr{reference image} remains constant, unlike \corr{the case of} running-difference images \corr{that use} a changing \corr{reference image.}

\corr{The pre-event image is used to define an area mask on the surface of the Sun bounded by two great-circles 10$^{\circ}$ apart that intersect at the location of the flare (shown in red in Figure~\ref{fig:wave_prof}a). Each mask is segmented into} a series of annuli of 1\corr{$^\circ$} width on the solar surface from the source point to the solar limb. This width was chosen due to the arbitrary nature of the wave propagation; the \corr{mask annuli} used to identify the pulse will by definition cross pixels at an angle, so the 1\corr{$^\circ$} width ensures that sufficient whole pixels will be included in each \corr{annulus} while negating the effects of pixel fragments at \corr{the} edges. \corr{This process is repeated to achieve 360 overlapping area masks sequentially offset by 1$^{\circ}$.} These \corr{masks} are unique to each event and are defined once and called from memory as required for that event. This approach speeds up a repetitive process and ensures that all images are treated equally and systematically.

\subsection{Image Processing}\label{subsect:process}

The pulse is identified using intensity profiles obtained from each \corr{area mask} for each successive image. This approach allows the position, pulse width, and maximum intensity to be determined for each observation\corr{. The pulse features being detected are typically broader than 10$^{\circ}$, but can exhibit spatial inhomogeneity on smaller scales. The area masks are therefore chosen to overlap to ensure that a smooth variation is found in the spatial location of the pulse.} 

The images are processed on an image-by-image basis, allowing for continuous data streaming and in recognition of the fact that each pulse is different and can last for a period of time that is \emph{a priori} unknown. Due to variations in the image intensity, exposures affected by the onboard Automatic Exposure Control (AEC) system are ignored by \textsf{CorPITA} as they adversely affect the PBD image intensity used to identify features. Each image is processed using the \textsf{aia\_prep.pro} routine available in the SolarSoftWare library and rebinned to $2048\times2048$~pixels before being de-rotated to the time of the base image and then combined with the base image to create a PBD image. This rebinning speeds up the processing of the image (a requirement for near-real-time detection) while ensuring sufficient pixels and retaining a high degree of accuracy.

\corr{An intensity profile is then created for each of the 360 overlapping area masks previously described in Section~\ref{subsect:prep}. The mean intensity and corresponding standard deviation is calculated from the set of pixels contained within each mask annulus. For a given area mask, this results in a 1-dimensional mean intensity profile that varies with distance away from the source location (e.g. Figure~\ref{fig:wave_prof}c). The pulse is subsequently} identified using a watershed technique and fitted using a Gaussian model \citep[cf.][]{Wills-Davey:1999fc}. 

Three peaks in intensity are identified at this point, allowing any stationary bright features or the flare to be identified and discarded if necessary and ensuring an accurate fit to the pulse. This is shown in Figure~\ref{fig:wave_prof}c where three intensity peaks have been successfully identified and fitted by the algorithm. The peak identification for each arc uses the parameters derived for the previous \corr{profile} as an initial estimate to ensure consistent results. This process is repeated for each \corr{profile} and the parameters saved.

The algorithm repeats this approach for three images before beginning to look for motion (three images are used as this is the minimum number of images required to produce an average motion, smoothing out image-to-image variation). The algorithm examines the distance--time variation of the features identified for each \corr{profile} using the derivative of a Savitsky--Golay filter. This determines forward motion by checking if the velocity of the identified feature is positive and lies within pre-defined limits (i.e. $1<v<2\,000$~km~s$^{-1}$). These limits ensure that a significant jump in feature position (e.g. from close to the source to close to the limb in a single time-step) is not mistaken for actual forward motion. This test is reapplied as new images are processed, allowing a continuous examination of whether the identified features are exhibiting forward motion.

This process is repeated for ten minutes worth of observations, allowing a significant amount of time for the pulse to appear and begin propagating (if present). This time period was chosen as it is comparable to the typical observing cadence of the \corr{\emph{Extreme UltraViolet Imager}} \citep[EUVI;][]{Wuelser:2004bd} onboard the \emph{Solar TErrestrial RElations Observatory} \citep[STEREO;][]{Kaiser:2008ij} spacecraft. If a pulse has been identified at this point moving away from the source and is still propagating, the algorithm continues operation on an image-by-image basis until \corr{continuous outward motion is no longer found}.

\subsection{Event Detection}\label{subsect:analysis}

\begin{figure*}[!t]
\begin{center}
\includegraphics[keepaspectratio,width=0.9\textwidth,trim=0 0 0 0]{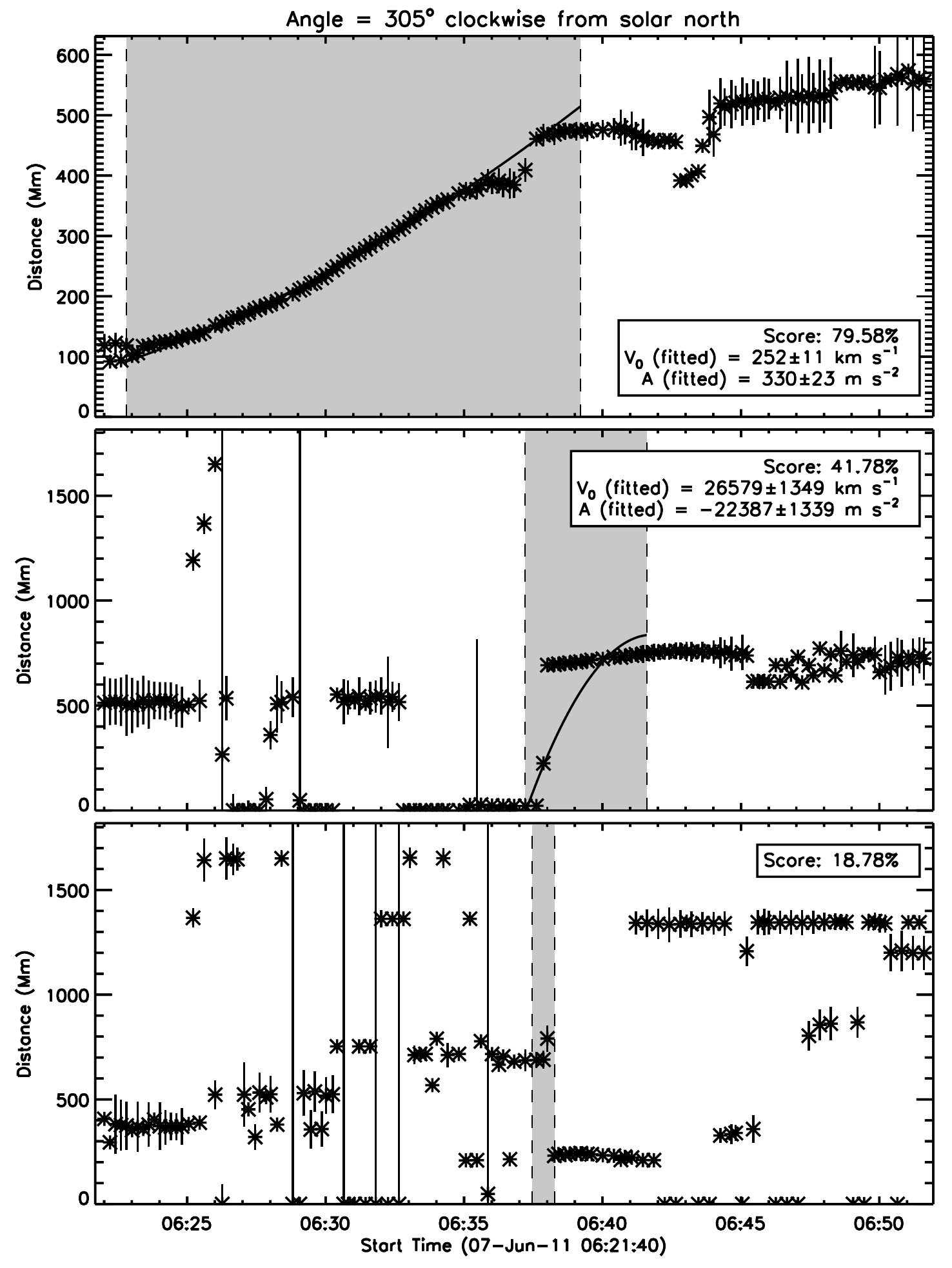}
\end{center}
\caption{Variation in distance with time for the three candidate features identified using the \corr{area mask} shown in Figure~\ref{fig:wave_prof}. \corr{The start time quoted on the horizontal axis indicates the time of the triggering event and the first image processed, with the base reference image obtained 120~s prior to this time.} The shaded areas indicate the sectors identified as exhibiting possible pulse propagation by \corr{\textsf{CorPITA}}. It is clear that candidate~1 (top) exhibits continuous outward motion within this section, and this is shown in the score of \corr{$\approx$}80~\%; whereas candidate~2 (middle) exhibits a significant jump between data points, while insufficient data points were identified for candidate~3 (bottom). This is reflected in the scores of \corr{$\approx$}42~\% and \corr{$\approx$}19~\% for candidates~2 and 3, respectively.}
\label{fig:wave_dt}
\end{figure*}

Once a moving feature has been identified, tracked\corr{,} and can no longer be detected, its properties are examined to determine if it is a propagating pulse or some other phenomenon (such \corr{as a} moving loop or filament eruption). The variation \corr{of feature} distance with time is examined for each of the three candidates identified for \corr{an area mask. The algorithm identifies} the \corr{longest time period} of continuous outward motion \corr{for each of the three candidates, scoring each} to determine \corr{that} which best corresponds to a pulse \corr{in that area mask.}

The scoring system uses a number of different parameters to try and identify a propagating ``EIT wave'' with the parameters designed to return a percentage ``quality-rating''. The most important parameter (accounting for 50\corr{\,\%} of the quality-rating score) is the number of data points corresponding to the detected pulse \corr{($N_{\mathrm{sector}}$) relative to} the total number of images processed \corr{($N_{\mathrm{total}}$)}. The other parameters are scored using a simple pass/fail system \corr{that awards} 1~point for a pass and 0~points for a fail\corr{. The first two of these parameters examine the kinematics of the pulse, with} points awarded for reasonable fit values of \corr{initial} velocity ($v_{\mathrm{fit}}$) and \corr{constant} acceleration ($a_{\mathrm{fit}}$)\corr{. The final parameter examines the uncertainty in the fitted pulse position, quantified by the relative position error, defined as,
\begin{equation}
\sigma^{\mathrm{rel}} = \mathrm{mean}\left(\frac{\sigma_d}{d}\right),
\end{equation}
where $d$ is an array of pulse fit-centroid distances from the source location, $\sigma_d$ is an array of fit uncertainties in $d$, and the mean is calculated over all data points in the longest period of continuous outward motion. The quality-rating scoring system then follows,
\begin{equation}
\textrm{score} = \left[\left(\frac{1}{2} \times \frac{N_{\mathrm{sector}}}{N_{\mathrm{total}}}\right) + \frac{(v_{\mathrm{score}} + a_{\mathrm{score}} + \sigma^{\mathrm{rel}}_{\mathrm{score}})}{6}\right] \times 100,
\end{equation}
where $v_\mathrm{score}$ has a value of 1 for $1<v_{fit}<2\,000$~km~s$^{-1}$, $a_{\mathrm{score}}$ has a value of 1 for $-2\,000<a_{\mathrm{fit}}<2\,000$~m~s$^{-2}$, and $\sigma^{\mathrm{rel}}_{\mathrm{score}}$ has a value of 1 for $\sigma^{\mathrm{rel}}<0.5$. }

Figure~\ref{fig:wave_dt} shows the variation in distance with time for the three candidates identified by \corr{\textsf{CorPITA}} for the \corr{area mask} shown in Figure~\ref{fig:wave_prof}. In each case, the shaded section indicates the possible pulse identified by the algorithm, with the quality rating then used to determine the optimal candidate for the ``EIT wave'' pulse. The top panel shows a candidate which exhibits continuous outward motion over an extended period of time with no major errors associated with the data points and reasonable estimates for the fitted kinematics. This is reflected in the quality rating score of \corr{$\approx$80\,\%}. The middle panel shows a candidate \corr{that} exhibits some variation in position with time where \corr{\textsf{CorPITA}} has identified a small section which may correspond to a pulse. However, a significant jump in distance over two data-points is apparent, producing unphysical estimates for the fitted kinematics, all of which is apparent in the small quality score. Finally, the third candidate exhibits significant variability in distance with time, and only a very small section \corr{that} may correspond to a pulse is identified. This section is too small to fit kinematics and this is reflected in the low quality score. 


Each candidate from each arc is scored, with the scores recorded and used to identify the presence of a pulse. If the quality score for the highest rated candidate from an arc exceeds 60\corr{\,\%}, a positive identification is recorded for that arc. A positive pulse identification is recorded only if \corr{ten} or more adjacent arcs record a score greater than 60\corr{\,\%}. This allows highly directional pulses to be identified, but also ensures that pulses are not inaccurately identified by multiple single arcs returning anomalously high scores. A series of plots recording the behaviour and kinematic properties of the candidates identified in each arc \corr{is} then produced allowing a visual confirmation if required.

\subsection{Characterisation and Output}\label{subsect:output}

\begin{figure*}[!t]
\begin{center}
\includegraphics[keepaspectratio,width=0.85\textwidth,trim=0 40 0 60]{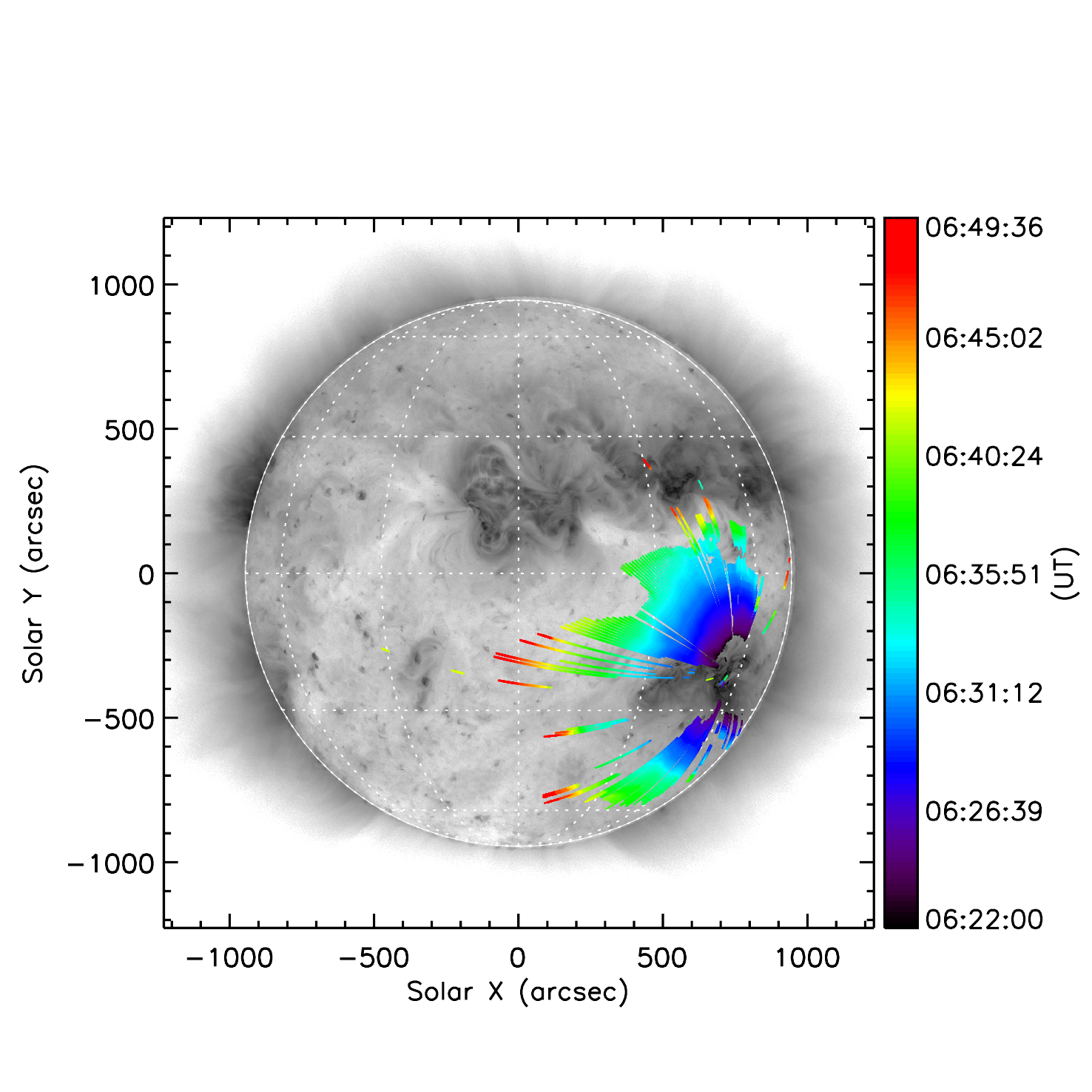}
\end{center}
\caption{The evolution of the pulse detected by \corr{\textsf{CorPITA}} for the 7~June~2011 eruption. The temporal variation of the pulse is indicated by colour, with the key shown in the colour-bar on the right-hand side.}
\label{fig:wave_img}
\end{figure*}

Once a pulse has been positively identified, \corr{\textsf{CorPITA}} returns a series of outputs allowing a thorough examination of the detection. The first output illustrates the evolution of the pulse as it propagates across the Sun; this is shown in Figure~\ref{fig:wave_img}. In this case, colour indicates the \corr{temporal} evolution of the pulse, with each \corr{area mask} that returns a sufficiently high-quality score being plotted. This allows a rapid assessment of the temporal evolution of the pulse, as well as allowing the behaviour of the pulse as it encounters different regions of the solar atmosphere to be highlighted.

\begin{figure*}[!t]
\begin{center}
\includegraphics[keepaspectratio,width=0.85\textwidth,trim=0 20 0 0]{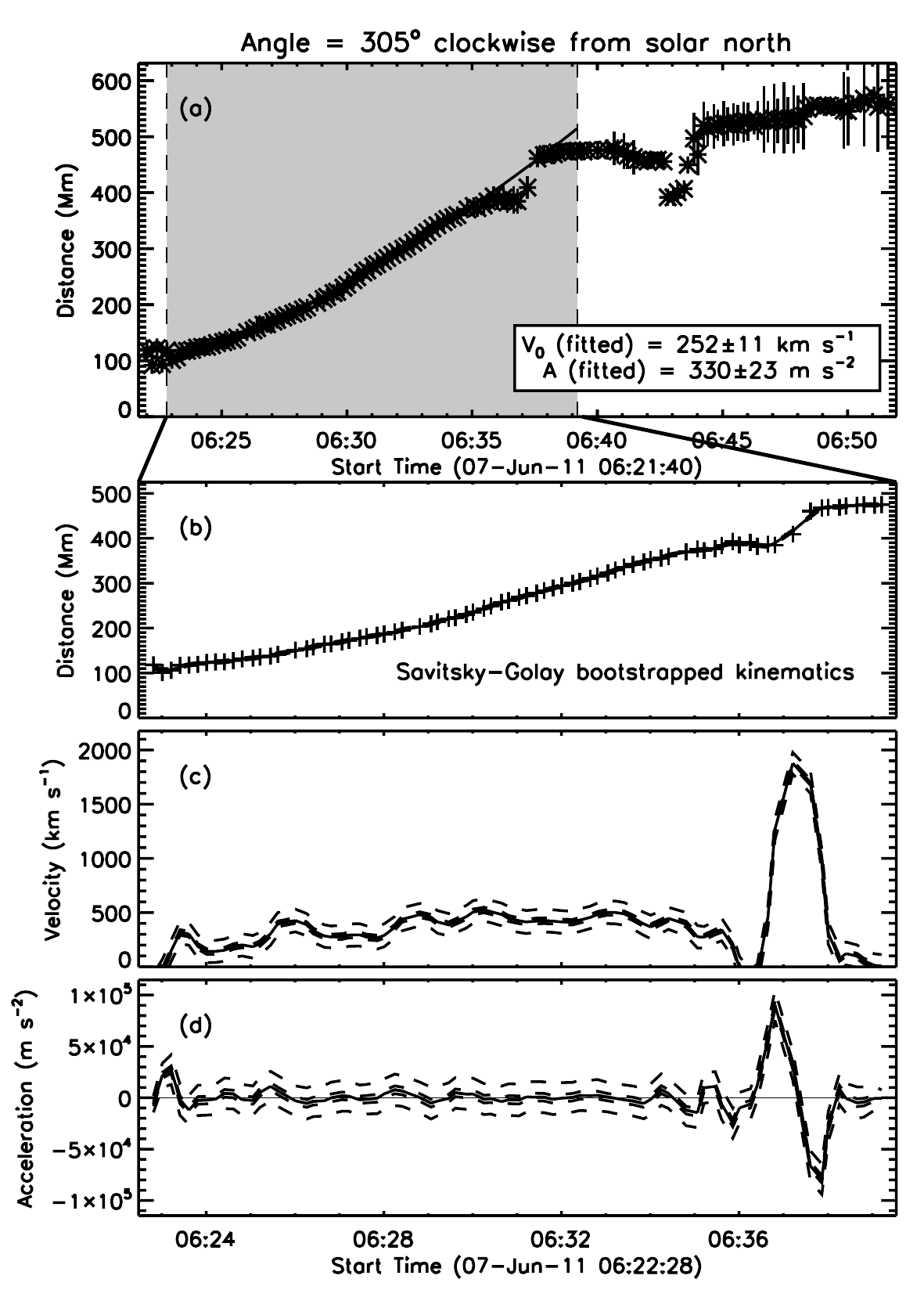}
\end{center}
\caption{The kinematics of the pulse for the \corr{area mask} at 305$^{\circ}$ clockwise from solar north (previously shown in Figure~\ref{fig:wave_prof} and corresponding to candidate~1 in Figure~\ref{fig:wave_dt}) derived using a quadratic fit (panel a with parameters quoted) and a Savitsky-Golay filter (panels b, c and d). The median (\corr{i.e., the 50th percentile of the data;} \emph{solid line}), \corr{the interquartile range boundaries} (\corr{i.e., the 25th and 75th percentiles of the data;} \emph{inner dashed lines}) and upper and lower fences (\emph{outer dashed lines}) have also been over-plotted on panels c and d.}
\label{fig:wave_kins}
\end{figure*}

The variation in derived kinematics with time for the identified pulse are also returned for each \corr{area mask} studied. As well as returning the initial velocity and constant acceleration values from a quadratic fit to the data, a bootstrapped Savitsky--Golay filter \citep[cf.][]{Byrne:2013kx} is used to determine the unbiased variation in both velocity and acceleration with time (see Figure~\ref{fig:wave_kins}). It is clear that while the fit to the distance--time data shown in Figure~\ref{fig:wave_kins}a returns a single value for the initial velocity and acceleration, the Savitsky--Golay filter reveals the variation in both velocity and acceleration during propagation, with the bootstrapping approach allowing the associated errors to be \corr{well} defined \corr{\citep[cf.][]{Byrne:2013kx}}. This approach may potentially be used to examine plasma variations in the material \corr{through which the pulse propagates} \citep[e.g.,][]{West:2011rq,Long:2011dz,Long:2013fu}. 

With a positive pulse identification, \corr{\textsf{CorPITA}} produces an output XML file for input into the HEK, recording the source location, maximum quality score for the event, start and end time for the pulse, and the fitted kinematics and direction of the highest-rated overall candidate to have been identified. This allows a brief summary of the event and provides a simple quotable estimate for the fitted kinematics of the pulse.


\section{Case Study Results}\label{sect:app}

\corr{\textsf{CorPITA}} was originally defined and tested using the eruption from 7~June~2011 as a sample case as it featured a spectacular filament eruption with a number of active regions and coronal holes on disk, which made identifying and tracking the propagating disturbance difficult. However, a systematic analysis of a larger data set from February~2011 is included here as a sample case study to illustrate the capabilities of the algorithm. This time period was chosen as a number of flares of varying size and magnitude were observed during that month, with NOAA~AR~11158 in particular being quite active. A number of ``EIT waves'' were also observed, several of which have been previously studied in detail \citep[see, e.g.,][]{Schrijver:2011qo,Veronig:2011mb,Harra:2011hc,Olmedo:2012ff,Long:2013fu}, allowing a direct comparison of results. 

A list of flares identified using the 94~\AA\ passband \corr{of} SDO/AIA by the SDO \corr{\textsf{Flare Detective Algorithm}} \citep{Grigis:2010tg} was used as a basis to define the location and start time of all flares during February~2011. Note that the flare location provided by the \corr{\textsf{Flare Detective}} is the centre of the macro-pixel used to identify the flare rather than the actual position of the flare itself. \corr{This may be compared to the issues with identifying a feature of finite initial extent erupting from anywhere in an irregularly shaped active region and generally not observed until propagating through the quiet corona. The use of an initial offset distance compensates for both the macro-pixel offset and the initial uncertainty with regard to the true eruption centre of the feature but} does not affect the ultimate identification or measurement of the pulse properties as \corr{it} is just chosen for consistency as a point \corr{from} which to measure position. 

\begin{figure*}[!t]
\begin{center}
\includegraphics[keepaspectratio,width=1\textwidth,trim=0 10 0 0]{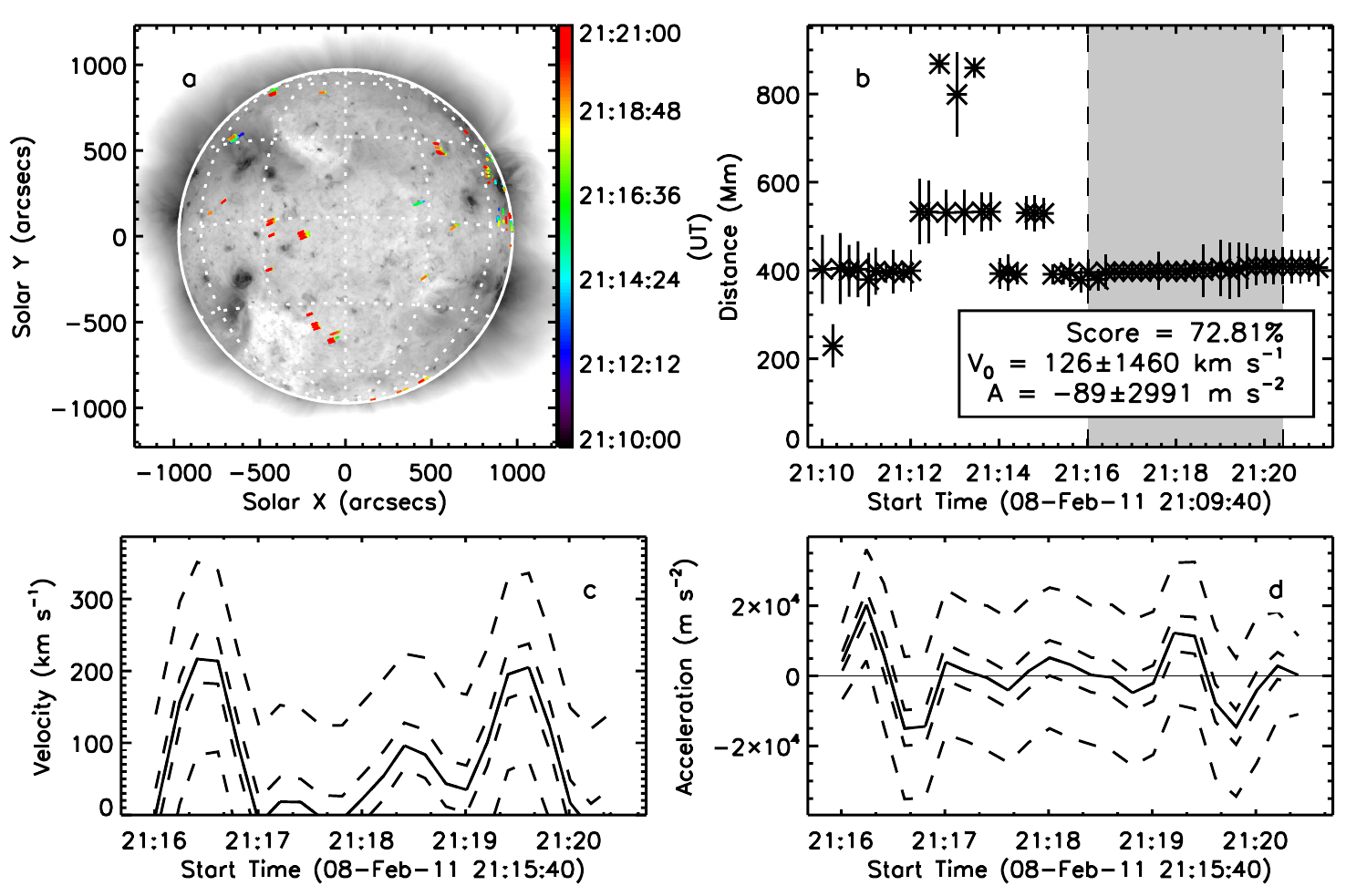}
\end{center}
\caption{Example of a flare from 8~February~2011 processed by CorPITA and returned as a non-wave. Panel a shows the map of overall pulse propagation; panel b shows the distance variation with time of the highest-rated pulse; panels c and d show the resulting Savitsky--Golay derived velocity and acceleration respectively. Although the highest-rated pulse scored $\approx$73~\% and 57 arcs scored more than 60~\%, these were \corr{physically dispersed across the disk} so this event was not identified as a positive wave detection.}
\label{fig:non_wave}
\end{figure*}

\corr{A series of 47} flares of varying magnitude were identified, including one X-class flare, seven M-class flares, \corr{24} C-class flares and \corr{15} flares with no measured GOES class. \corr{\textsf{CorPITA}} was applied to each event in turn, with the data automatically downloaded as required from the Virtual Solar Observatory. Of the \corr{47} flares identified, \corr{17} were associated with a wave as detected by \corr{\textsf{CorPITA}}, including the one X-class flare, five of the M-class flares, eight of the C-class flares and three with no measured GOES class. \corr{No wave signature was observed for 24} of the identified and tested flares. 

The robust nature of the algorithm is illustrated by Figure~\ref{fig:non_wave}, which shows an event from 8~February~2011 which had no detectable wave signature. The highest rated feature detected (shown in Figure~\ref{fig:non_wave}b\,--\,d) has a score of \corr{$\approx$}73~\%, with \corr{57} arcs exhibiting detected features with a score greater than 60~\%. However, these arcs were randomly oriented and did not fulfil the criterion of at least ten adjacent arcs with a score greater than 60~\%, with the result that the event was correctly classified as a ``non-wave'' event. The random nature of the detections are apparent in Figure~\ref{fig:non_wave}a, where the speckled dots indicate detections. 

\begin{figure*}[!t]
\begin{center}
\includegraphics[keepaspectratio,width=1\textwidth,trim=0 10 0 0]{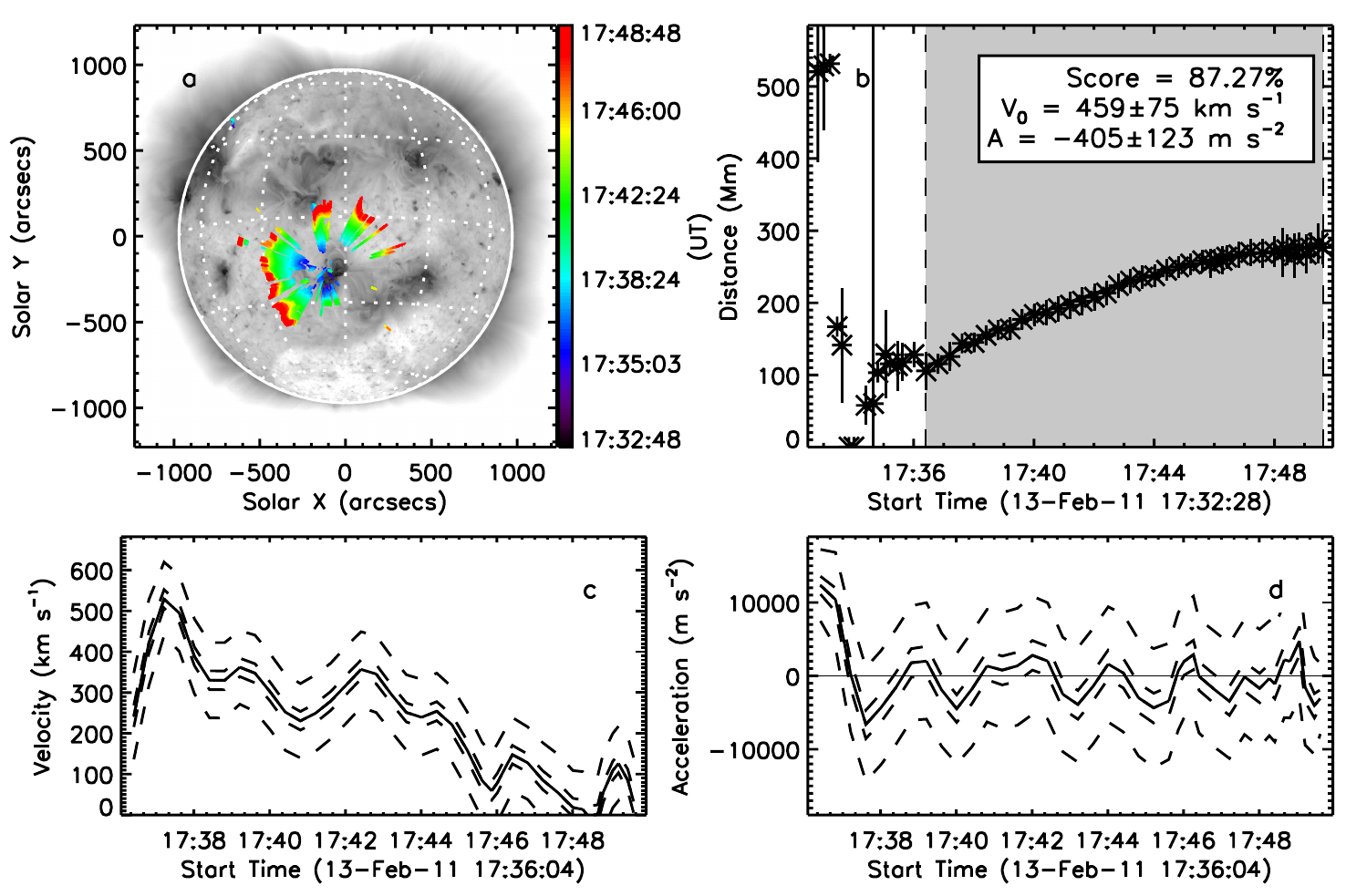}
\end{center}
\caption{\corr{Example of the \textsf{CorPITA} output for} a flare from 13~February~2011 identified as a wave by \corr{\textsf{CorPITA}}. Panel a shows the pulse propagation with time while panels b\,--\,d show the fitted kinematics and Savitsky--Golay derived velocity and acceleration, respectively, for the highest-rated \corr{area mask}.}
\label{fig:wave}
\end{figure*}

An example of a positive ``EIT wave'' detection identified by \corr{\textsf{CorPITA}} is shown in Figure~\ref{fig:wave} for an event \corr{that} erupted from NOAA~AR~11158 on 13~February~2011. The large-scale nature of the pulse is immediately apparent from Figure~\ref{fig:wave}a as it propagates in almost all directions from its epicentre. In addition, the epicentre of the pulse is offset from the centre of the active region, suggesting that the initial driver in this case erupted from the edge of the active region rather than the centre. The score of the highest rated \corr{area mask} for this event is $\approx$87~\%, only slightly higher than that shown in Figure~\ref{fig:non_wave}. However, the global nature of the feature detected in this case has been identified by the algorithm that correctly classified this event as an ``EIT wave''. 

The capabilities of the algorithm in detecting variations in the kinematics of the pulse as it propagates with time are also apparent in Figure~\ref{fig:wave}a. Although the pulse propagates in almost all directions from the epicentre, this propagation is not uniform. This has potential implications for coronal seismology by allowing ``EIT waves'' to be used to study the properties of the solar corona on a systematic basis. The combination of pulse detections from \corr{\textsf{CorPITA}} with one of either global magnetic-field extrapolations or density estimates could be used to make global estimates of the other. In addition, regions that are particularly ``EIT wave'' productive would potentially allow estimates of plasma properties to be made over extended periods of time (i.e., an entire disk passage).

\section{Discussion and Conclusions} \label{sect:conc} 

In this \corr{article} we have presented a new automated algorithm for identifying, tracking and analysing ``EIT waves'' in data from SDO/AIA. \corr{\textsf{CorPITA}} uses an intensity-profile technique applied to percentage base-difference images to identify the propagating pulse, tracking it for as long as possible before estimating the variation in kinematics across 360 overlapping \corr{area masks} of 10$^\circ$ width. This allows the variation of the pulse propagation across the entire Sun to be studied and analysed, providing an indication of variations in the magnetic-field strength and density of the low solar corona. The systematic, automated approach of \corr{\textsf{CorPITA} allows reproducible detections of an ``EIT wave'' following a specified series of analytical steps along with a statistically significant} estimate of the pulse kinematics, providing an opportunity to determine their true physical nature across a large sample of events.

\begin{figure*}[!t]
\begin{center}
\includegraphics[keepaspectratio,width=0.97\textwidth,trim=0 0 0 0]{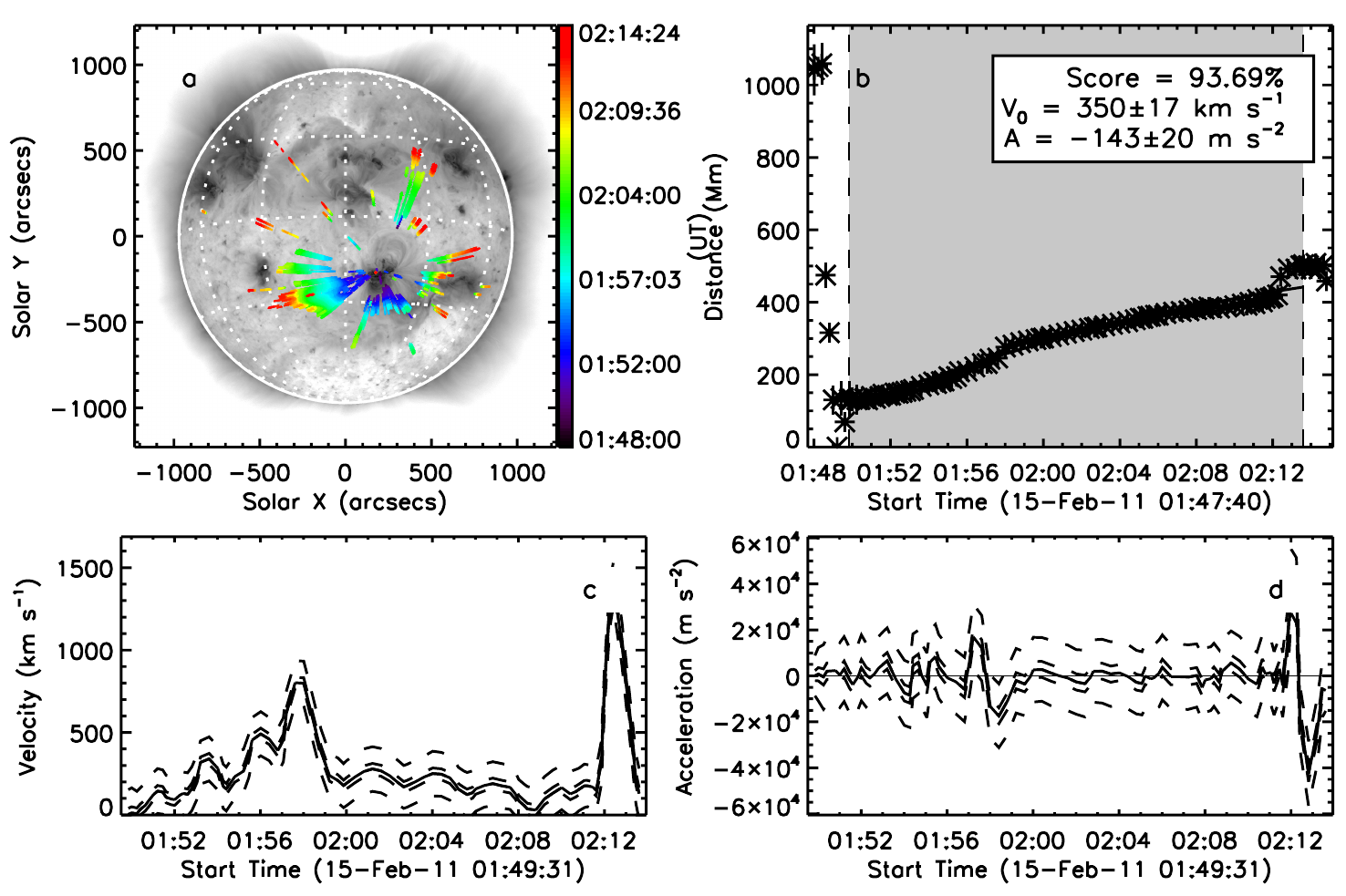}
\includegraphics[keepaspectratio,width=0.97\textwidth,trim=0 10 0 0]{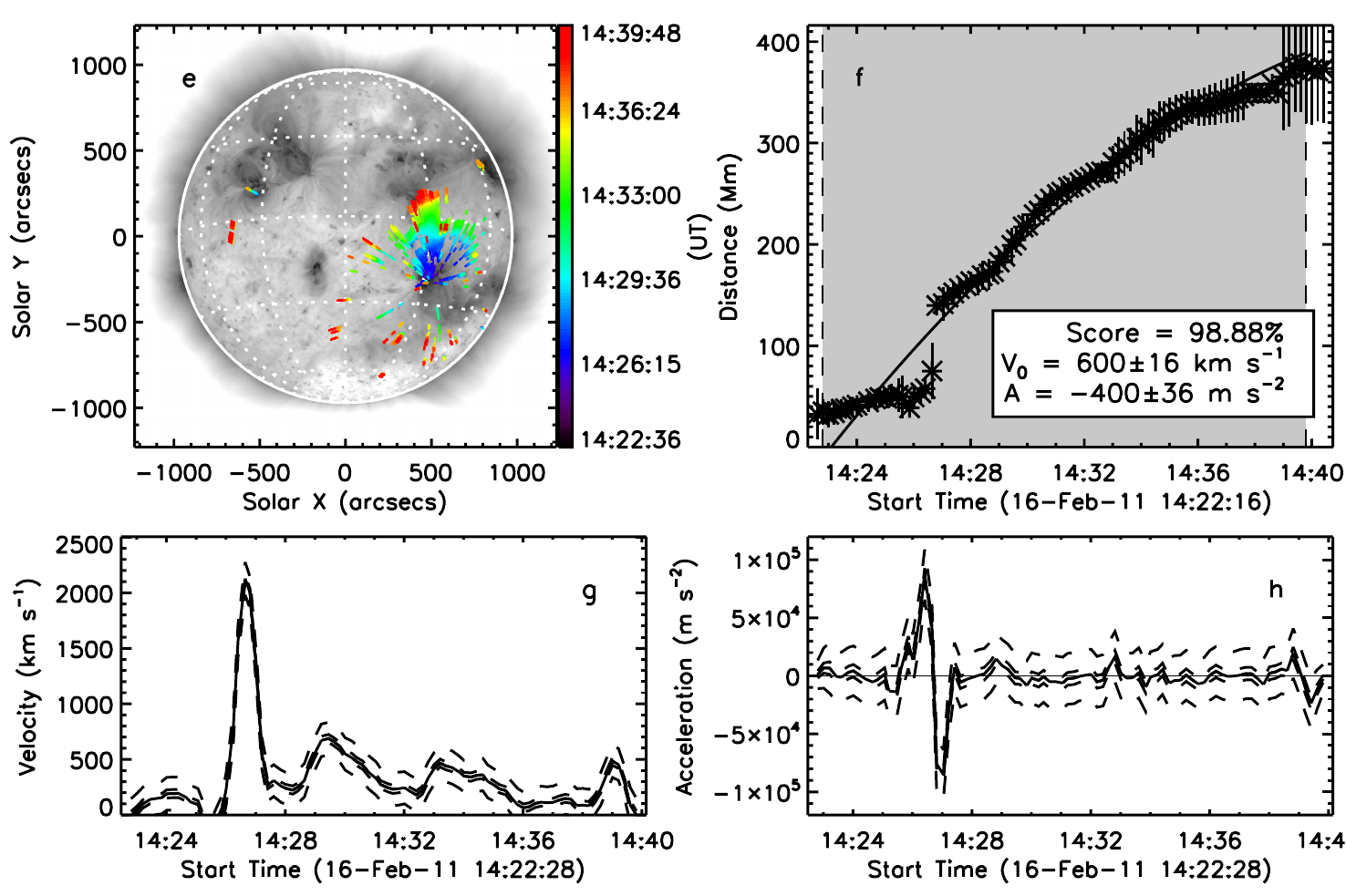}
\end{center}
\caption{\corr{Example of the \textsf{CorPITA} output} for the events from 15~February~2011 \citep[panels a\,--\,d, previously studied by][]{Schrijver:2011qo,Olmedo:2012ff} and 16~February~2011 \citep[panels e\,--\,h, previously studied by][]{Harra:2011hc,Veronig:2011mb,Long:2013fu}. \corr{For the 15~February~2011 (16~February~2011) event, panel a (e) shows the pulse propagation with time while panels b\,--\,d (f\,--h) show the fitted kinematics and Savitsky--Golay derived velocity and acceleration, respectively, for the highest-rated arc. \textsf{CorPITA}} has successfully identified ``EIT waves'' in both cases indicating that the algorithm is robust.}
\label{fig:prev_wave}
\end{figure*}

The period chosen to provide a sample output from \corr{\textsf{CorPITA}} included two large-scale ``EIT wave'' events that have been previously studied. The \corr{\textsf{CorPITA}} analysis of both of these events is shown in Figure~\ref{fig:prev_wave}. In both cases, ``EIT waves'' were successfully identified by the algorithm, allowing a direct comparison with the work of \citet{Schrijver:2011qo} and \citet{Olmedo:2012ff}, who studied the 15~February~2011 event, and \citet{Harra:2011hc}, \citet{Veronig:2011mb} and \citet{Long:2013fu}, who studied the 16~February~2011 event. 

For the 15~February~2011 event (shown in Figure~\ref{fig:prev_wave}a\,--\,d), \corr{\textsf{CorPITA}} returned an average initial velocity of $406\pm1$~km~s$^{-1}$, although this varied from $\approx$\,0\,--\,1\,000~km~s$^{-1}$ across the arcs where a pulse was detected and tracked. Figure~\ref{fig:prev_wave}a also shows the anisotropic nature of the pulse propagation, with the pulse tracked primarily across quiet regions of the solar corona and not through the different adjacent active regions. This is consistent with the work of \citet{Olmedo:2012ff} who noted the variation in pulse velocity with direction, although the current version of \corr{\textsf{CorPITA}} cannot detect the reflection that they observed while the transmission through coronal holes and active regions would produce a jump in pulse position that cannot currently be tracked. Despite this, \corr{\textsf{CorPITA}} successfully identifies the pulse and the variation in its propagation through the quiet solar corona.

The 16~February~2011 event (shown in Figure~\ref{fig:prev_wave}e\,--\,h) was also successfully identified by \corr{\textsf{CorPITA}}, with an average initial velocity of $331\pm6$~km~s$^{-1}$ and fitted initial velocities varying from $\approx$\,100\,-\,975~km~s$^{-1}$ across the different arcs. The anisotropic nature of the pulse is again shown by Figure~\ref{fig:prev_wave}e, with the pulse in this case strongly directed towards solar North away from the erupting active region. The fitted initial velocities measured by \corr{\textsf{CorPITA}} are in good correspondence with those measured by \citet{Harra:2011hc}, \citet{Veronig:2011mb} and \citet{Long:2013fu}, indicating that the same pulse is identified. However, \corr{\textsf{CorPITA} is designed to identify the forward motion of a single propagating pulse and} cannot yet identify multiple \corr{propagating} pulses associated with a single event\corr{, although this may be implemented in a future iteration of the code. As a result, it} does not observe the second propagating feature identified by \citet{Harra:2011hc}.

The results shown here indicate that \corr{\textsf{CorPITA}} offers a fully automated, robust approach for identifying, tracking and analysing ``EIT waves''. This will allow \corr{a systematically reproducible} analysis of the entire SDO/AIA data-set as well as a near-real-time analysis of these events when fully incorporated into the SDO feature analysis pipeline. Automating the identification of these events from SDO/AIA data will allow an improved statistical analysis which has implications for our understanding of the physical processes involved in the eruption and propagation of this phenomenon. The ability of \corr{\textsf{CorPITA}} to identify anisotropies in the propagation of these pulses also suggests that it may be used to investigate the large-scale structure of the solar corona using coronal seismology to study density and magnetic field variations.

\begin{acks}
The authors wish to thank Rebecca Feeney-Barry for useful discussions\corr{, and the anonymous referee whose comments helped to improve the paper}. The SDO feature finding team effort is supported by NASA. Data from SDO/AIA are courtesy of NASA/SDO and the AIA science team. DML received funding from the European Commission's Seventh Framework Programme under the grant agreement No. 284461 (eHEROES project), while DSB was funded by the European Space Agency Prodex programme.
\end{acks}

\bibliographystyle{spr-mp-sola}
\bibliography{bibtex_lib.bib}  

\begin{thebibliography}{48}
\ifx \bisbn   \undefined \def \bisbn  #1{ISBN #1}\fi
\ifx \binits  \undefined \def \binits#1{#1}\fi
\ifx \bauthor  \undefined \def \bauthor#1{#1}\fi
\ifx \batitle  \undefined \def \batitle#1{#1}\fi
\ifx \bjtitle  \undefined \def \bjtitle#1{\textit{#1}}\fi
\ifx \bvolume  \undefined \def \bvolume#1{\textbf{#1}}\fi
\ifx \byear  \undefined \def \byear#1{#1}\fi
\ifx \bissue  \undefined \def \bissue#1{#1}\fi
\ifx \bfpage  \undefined \def \bfpage#1{#1}\fi
\ifx \blpage  \undefined \def \blpage #1{#1}\fi
\ifx \burl  \undefined \def \burl#1{\textsf{#1}}\fi
\ifx \href  \undefined \def \href#1#2{\textsf{#2}}\fi
\ifx \doiurl  \undefined \def
  \doiurl#1{\href{http://dx.doi.org/#1}{\textsf{#1}}}\fi
\ifx \betal  \undefined \def \betal{\textit{et al.}}\fi
\ifx \binstitute  \undefined \def \binstitute#1{#1}\fi
\ifx \bctitle  \undefined \def \bctitle#1{#1}\fi
\ifx \beditor  \undefined \def \beditor#1{#1}\fi
\ifx \bpublisher  \undefined \def \bpublisher#1{#1}\fi
\ifx \bbtitle  \undefined \def \bbtitle#1{\textit{#1}}\fi
\ifx \bedition  \undefined \def \bedition#1{#1}\fi
\ifx \bseriesno  \undefined \def \bseriesno#1{\textbf{#1}}\fi
\ifx \blocation  \undefined \def \blocation#1{#1}\fi
\ifx \bsertitle  \undefined \def \bsertitle#1{\textit{#1}}\fi
\ifx \bsnm \undefined \def \bsnm#1{#1}\fi
\ifx \bsuffix \undefined \def \bsuffix#1{#1}\fi
\ifx \bparticle \undefined \def \bparticle#1{#1}\fi
\ifx \barticle \undefined \def \barticle#1{}\fi
\ifx \botherref \undefined \def \botherref#1{}\fi
\ifx \url \undefined \def \url#1{\textsf{#1}}\fi
\ifx \bchapter \undefined \def \bchapter#1{}\fi
\ifx \bbook \undefined \def \bbook#1{}\fi
\ifx \bcomment \undefined \def \bcomment#1{#1}\fi
\ifx \oauthor \undefined \def \oauthor#1{#1}\fi
\ifx \citeauthoryear \undefined \def \citeauthoryear#1{#1}\fi
\def \endbibitem {}
\ifx \bconflocation  \undefined \def \bconflocation#1{#1} \fi

\bibitem[\protect\citeauthoryear{{Aschwanden}}{2010}]{Aschwanden:2010fk}
\begin{barticle}
\bauthor{\bsnm{{Aschwanden}}, \binits{M.J.}}:
\byear{2010},
\batitle{{Image Processing Techniques and Feature Recognition in Solar
  Physics}}.
\bjtitle{\solphys}
\bvolume{262},
\bfpage{235}\,--\,\blpage{275}.
doi:\doiurl{10.1007/s11207-009-9474-y}.
\end{barticle}
\endbibitem

\bibitem[\protect\citeauthoryear{{Biesecker}
  \textit{et~al.}}{2002}]{Biesecker:2002uq}
\begin{barticle}
\bauthor{\bsnm{{Biesecker}}, \binits{D.A.}},
\bauthor{\bsnm{{Myers}}, \binits{D.C.}},
\bauthor{\bsnm{{Thompson}}, \binits{B.J.}},
\bauthor{\bsnm{{Hammer}}, \binits{D.M.}},
\bauthor{\bsnm{{Vourlidas}}, \binits{A.}}:
\byear{2002},
\batitle{{Solar Phenomena Associated with ``EIT Waves''}}.
\bjtitle{\apj}
\bvolume{569},
\bfpage{1009}\,--\,\blpage{1015}.
doi:\doiurl{10.1086/339402}.
\end{barticle}
\endbibitem

\bibitem[\protect\citeauthoryear{{Byrne} \textit{et~al.}}{2013}]{Byrne:2013kx}
\begin{barticle}
\bauthor{\bsnm{{Byrne}}, \binits{J.P.}},
\bauthor{\bsnm{{Long}}, \binits{D.M.}},
\bauthor{\bsnm{{Gallagher}}, \binits{P.T.}},
\bauthor{\bsnm{{Bloomfield}}, \binits{D.S.}},
\bauthor{\bsnm{{Maloney}}, \binits{S.A.}},
\bauthor{\bsnm{{McAteer}}, \binits{R.T.J.}},
\bauthor{\bsnm{{Morgan}}, \binits{H.}},
\bauthor{\bsnm{{Habbal}}, \binits{S.R.}}:
\byear{2013},
\batitle{{Improved methods for determining the kinematics of coronal mass
  ejections and coronal waves}}.
\bjtitle{\aap}
\bvolume{557},
\bfpage{A96}.
doi:\doiurl{10.1051/0004-6361/201321223}.
\end{barticle}
\endbibitem

\bibitem[\protect\citeauthoryear{{Chen} and {Shibata}}{2002}]{Chen:2002zr}
\begin{bchapter}
\bauthor{\bsnm{{Chen}}, \binits{P.F.}},
\bauthor{\bsnm{{Shibata}}, \binits{K.}}:
\byear{2002},
\bctitle{{A Further Consideration of the Mechanism for EIT Waves}}.
In: \beditor{\bsnm{{Ikeuchi}}, \binits{S.}},
\beditor{\bsnm{{Hearnshaw}}, \binits{J.}},
\beditor{\bsnm{{Hanawa}}, \binits{T.}} (eds.)
\bbtitle{8th Asian-Pacific Regional Meeting}
\bseriesno{II},
\bfpage{421}\,--\,\blpage{422}.
\end{bchapter}
\endbibitem

\bibitem[\protect\citeauthoryear{{Chen} and {Wu}}{2011}]{Chen:2011vn}
\begin{barticle}
\bauthor{\bsnm{{Chen}}, \binits{P.F.}},
\bauthor{\bsnm{{Wu}}, \binits{Y.}}:
\byear{2011},
\batitle{{First Evidence of Coexisting EIT Wave and Coronal Moreton Wave from
  SDO/AIA Observations}}.
\bjtitle{\apjl}
\bvolume{732},
\bfpage{L20}.
doi:\doiurl{10.1088/2041-8205/732/2/L20}.
\end{barticle}
\endbibitem

\bibitem[\protect\citeauthoryear{{Chen}, {Fang}, and
  {Shibata}}{2005}]{Chen:2005ys}
\begin{barticle}
\bauthor{\bsnm{{Chen}}, \binits{P.F.}},
\bauthor{\bsnm{{Fang}}, \binits{C.}},
\bauthor{\bsnm{{Shibata}}, \binits{K.}}:
\byear{2005},
\batitle{{A Full View of EIT Waves}}.
\bjtitle{\apj}
\bvolume{622},
\bfpage{1202}\,--\,\blpage{1210}.
doi:\doiurl{10.1086/428084}.
\end{barticle}
\endbibitem

\bibitem[\protect\citeauthoryear{{Cohen} \textit{et~al.}}{2009}]{Cohen:2009ly}
\begin{barticle}
\bauthor{\bsnm{{Cohen}}, \binits{O.}},
\bauthor{\bsnm{{Attrill}}, \binits{G.D.R.}},
\bauthor{\bsnm{{Manchester}}, \binits{W.B.} \bsuffix{IV}},
\bauthor{\bsnm{{Wills-Davey}}, \binits{M.J.}}:
\byear{2009},
\batitle{{Numerical Simulation of an EUV Coronal Wave Based on the 2009
  February 13 CME Event Observed by STEREO}}.
\bjtitle{\apj}
\bvolume{705},
\bfpage{587}\,--\,\blpage{602}.
doi:\doiurl{10.1088/0004-637X/705/1/587}.
\end{barticle}
\endbibitem

\bibitem[\protect\citeauthoryear{{Delaboudini{\`e}re}
  \textit{et~al.}}{1995}]{Delaboudiniere:1995ve}
\begin{barticle}
\bauthor{\bsnm{{Delaboudini{\`e}re}}, \binits{J.-P.}},
\bauthor{\bsnm{{Artzner}}, \binits{G.E.}},
\bauthor{\bsnm{{Brunaud}}, \binits{J.}},
\bauthor{\bsnm{{Gabriel}}, \binits{A.H.}},
\bauthor{\bsnm{{Hochedez}}, \binits{J.F.}},
\bauthor{\bsnm{{Millier}}, \binits{F.}},
\bauthor{\bsnm{{Song}}, \binits{X.Y.}},
\bauthor{\bsnm{{Au}}, \binits{B.}},
\bauthor{\bsnm{{Dere}}, \binits{K.P.}},
\bauthor{\bsnm{{Howard}}, \binits{R.A.}},
\bauthor{\bsnm{{Kreplin}}, \binits{R.}},
\bauthor{\bsnm{{Michels}}, \binits{D.J.}},
\bauthor{\bsnm{{Moses}}, \binits{J.D.}},
\bauthor{\bsnm{{Defise}}, \binits{J.M.}},
\bauthor{\bsnm{{Jamar}}, \binits{C.}},
\bauthor{\bsnm{{Rochus}}, \binits{P.}},
\bauthor{\bsnm{{Chauvineau}}, \binits{J.P.}},
\bauthor{\bsnm{{Marioge}}, \binits{J.P.}},
\bauthor{\bsnm{{Catura}}, \binits{R.C.}},
\bauthor{\bsnm{{Lemen}}, \binits{J.R.}},
\bauthor{\bsnm{{Shing}}, \binits{L.}},
\bauthor{\bsnm{{Stern}}, \binits{R.A.}},
\bauthor{\bsnm{{Gurman}}, \binits{J.B.}},
\bauthor{\bsnm{{Neupert}}, \binits{W.M.}},
\bauthor{\bsnm{{Maucherat}}, \binits{A.}},
\bauthor{\bsnm{{Clette}}, \binits{F.}},
\bauthor{\bsnm{{Cugnon}}, \binits{P.}},
\bauthor{\bsnm{{van Dessel}}, \binits{E.L.}}:
\byear{1995},
\batitle{{EIT: Extreme-Ultraviolet Imaging Telescope for the SOHO Mission}}.
\bjtitle{\solphys}
\bvolume{162},
\bfpage{291}\,--\,\blpage{312}.
doi:\doiurl{10.1007/BF00733432}.
\end{barticle}
\endbibitem

\bibitem[\protect\citeauthoryear{{Delann{\'e}e}}{2000}]{Delannee:2000bh}
\begin{barticle}
\bauthor{\bsnm{{Delann{\'e}e}}, \binits{C.}}:
\byear{2000},
\batitle{{Another View of the EIT Wave Phenomenon}}.
\bjtitle{\apj}
\bvolume{545},
\bfpage{512}\,--\,\blpage{523}.
doi:\doiurl{10.1086/317777}.
\end{barticle}
\endbibitem

\bibitem[\protect\citeauthoryear{{Delann{\'e}e}
  \textit{et~al.}}{2008}]{Delannee:2008qf}
\begin{barticle}
\bauthor{\bsnm{{Delann{\'e}e}}, \binits{C.}},
\bauthor{\bsnm{{T{\"o}r{\"o}k}}, \binits{T.}},
\bauthor{\bsnm{{Aulanier}}, \binits{G.}},
\bauthor{\bsnm{{Hochedez}}, \binits{J.-F.}}:
\byear{2008},
\batitle{{A New Model for Propagating Parts of EIT Waves: A Current Shell in a
  CME}}.
\bjtitle{\solphys}
\bvolume{247},
\bfpage{123}\,--\,\blpage{150}.
doi:\doiurl{10.1007/s11207-007-9085-4}.
\end{barticle}
\endbibitem

\bibitem[\protect\citeauthoryear{{Dere} \textit{et~al.}}{1997}]{Dere:1997fk}
\begin{barticle}
\bauthor{\bsnm{{Dere}}, \binits{K.P.}},
\bauthor{\bsnm{{Brueckner}}, \binits{G.E.}},
\bauthor{\bsnm{{Howard}}, \binits{R.A.}},
\bauthor{\bsnm{{Koomen}}, \binits{M.J.}},
\bauthor{\bsnm{{Korendyke}}, \binits{C.M.}},
\bauthor{\bsnm{{Kreplin}}, \binits{R.W.}},
\bauthor{\bsnm{{Michels}}, \binits{D.J.}},
\bauthor{\bsnm{{Moses}}, \binits{J.D.}},
\bauthor{\bsnm{{Moulton}}, \binits{N.E.}},
\bauthor{\bsnm{{Socker}}, \binits{D.G.}},
\bauthor{\bsnm{{St.~Cyr}}, \binits{O.C.}},
\bauthor{\bsnm{{Delaboudini{\`e}re}}, \binits{J.P.}},
\bauthor{\bsnm{{Artzner}}, \binits{G.E.}},
\bauthor{\bsnm{{Brunaud}}, \binits{J.}},
\bauthor{\bsnm{{Gabriel}}, \binits{A.H.}},
\bauthor{\bsnm{{Hochedez}}, \binits{J.F.}},
\bauthor{\bsnm{{Millier}}, \binits{F.}},
\bauthor{\bsnm{{Song}}, \binits{X.Y.}},
\bauthor{\bsnm{{Chauvineau}}, \binits{J.P.}},
\bauthor{\bsnm{{Marioge}}, \binits{J.P.}},
\bauthor{\bsnm{{Defise}}, \binits{J.M.}},
\bauthor{\bsnm{{Jamar}}, \binits{C.}},
\bauthor{\bsnm{{Rochus}}, \binits{P.}},
\bauthor{\bsnm{{Catura}}, \binits{R.C.}},
\bauthor{\bsnm{{Lemen}}, \binits{J.R.}},
\bauthor{\bsnm{{Gurman}}, \binits{J.B.}},
\bauthor{\bsnm{{Neupert}}, \binits{W.}},
\bauthor{\bsnm{{Clette}}, \binits{F.}},
\bauthor{\bsnm{{Cugnon}}, \binits{P.}},
\bauthor{\bsnm{{van Dessel}}, \binits{E.L.}},
\bauthor{\bsnm{{Lamy}}, \binits{P.L.}},
\bauthor{\bsnm{{Llebaria}}, \binits{A.}},
\bauthor{\bsnm{{Schwenn}}, \binits{R.}},
\bauthor{\bsnm{{Simnett}}, \binits{G.M.}}:
\byear{1997},
\batitle{{EIT and LASCO Observations of the Initiation of a Coronal Mass
  Ejection}}.
\bjtitle{\solphys}
\bvolume{175},
\bfpage{601}\,--\,\blpage{612}.
doi:\doiurl{10.1023/A:1004907307376}.
\end{barticle}
\endbibitem

\bibitem[\protect\citeauthoryear{{Domingo}, {Fleck}, and
  {Poland}}{1995}]{Domingo:1995dq}
\begin{barticle}
\bauthor{\bsnm{{Domingo}}, \binits{V.}},
\bauthor{\bsnm{{Fleck}}, \binits{B.}},
\bauthor{\bsnm{{Poland}}, \binits{A.I.}}:
\byear{1995},
\batitle{{The SOHO Mission: an Overview}}.
\bjtitle{\solphys}
\bvolume{162},
\bfpage{1}\,--\,\blpage{37}.
doi:\doiurl{10.1007/BF00733425}.
\end{barticle}
\endbibitem

\bibitem[\protect\citeauthoryear{{Downs} \textit{et~al.}}{2011}]{Downs:2011nx}
\begin{barticle}
\bauthor{\bsnm{{Downs}}, \binits{C.}},
\bauthor{\bsnm{{Roussev}}, \binits{I.I.}},
\bauthor{\bsnm{{van der Holst}}, \binits{B.}},
\bauthor{\bsnm{{Lugaz}}, \binits{N.}},
\bauthor{\bsnm{{Sokolov}}, \binits{I.V.}},
\bauthor{\bsnm{{Gombosi}}, \binits{T.I.}}:
\byear{2011},
\batitle{{Studying Extreme Ultraviolet Wave Transients with a Digital
  Laboratory: Direct Comparison of Extreme Ultraviolet Wave Observations to
  Global Magnetohydrodynamic Simulations}}.
\bjtitle{\apj}
\bvolume{728},
\bfpage{2}.
doi:\doiurl{10.1088/0004-637X/728/1/2}.
\end{barticle}
\endbibitem

\bibitem[\protect\citeauthoryear{{Downs} \textit{et~al.}}{2012}]{Downs:2012cr}
\begin{barticle}
\bauthor{\bsnm{{Downs}}, \binits{C.}},
\bauthor{\bsnm{{Roussev}}, \binits{I.I.}},
\bauthor{\bsnm{{van der Holst}}, \binits{B.}},
\bauthor{\bsnm{{Lugaz}}, \binits{N.}},
\bauthor{\bsnm{{Sokolov}}, \binits{I.V.}}:
\byear{2012},
\batitle{{Understanding SDO/AIA Observations of the 2010 June 13 EUV Wave
  Event: Direct Insight from a Global Thermodynamic MHD Simulation}}.
\bjtitle{\apj}
\bvolume{750},
\bfpage{134}.
doi:\doiurl{10.1088/0004-637X/750/2/134}.
\end{barticle}
\endbibitem

\bibitem[\protect\citeauthoryear{{Gallagher} and
  {Long}}{2011}]{Gallagher:2011oq}
\begin{barticle}
\bauthor{\bsnm{{Gallagher}}, \binits{P.T.}},
\bauthor{\bsnm{{Long}}, \binits{D.M.}}:
\byear{2011},
\batitle{{Large-scale Bright Fronts in the Solar Corona: A Review of ``EIT
  waves''}}.
\bjtitle{\ssr}
\bvolume{158},
\bfpage{365}\,--\,\blpage{396}.
doi:\doiurl{10.1007/s11214-010-9710-7}.
\end{barticle}
\endbibitem

\bibitem[\protect\citeauthoryear{{Gopalswamy}
  \textit{et~al.}}{2009}]{Gopalswamy:2009kl}
\begin{barticle}
\bauthor{\bsnm{{Gopalswamy}}, \binits{N.}},
\bauthor{\bsnm{{Yashiro}}, \binits{S.}},
\bauthor{\bsnm{{Temmer}}, \binits{M.}},
\bauthor{\bsnm{{Davila}}, \binits{J.}},
\bauthor{\bsnm{{Thompson}}, \binits{W.T.}},
\bauthor{\bsnm{{Jones}}, \binits{S.}},
\bauthor{\bsnm{{McAteer}}, \binits{R.T.J.}},
\bauthor{\bsnm{{Wuelser}}, \binits{J.-P.}},
\bauthor{\bsnm{{Freeland}}, \binits{S.}},
\bauthor{\bsnm{{Howard}}, \binits{R.A.}}:
\byear{2009},
\batitle{{EUV Wave Reflection from a Coronal Hole}}.
\bjtitle{\apjl}
\bvolume{691},
\bfpage{L123}\,--\,\blpage{L127}.
doi:\doiurl{10.1088/0004-637X/691/2/L123}.
\end{barticle}
\endbibitem

\bibitem[\protect\citeauthoryear{{Grigis}
  \textit{et~al.}}{2010}]{Grigis:2010tg}
\begin{bchapter}
\bauthor{\bsnm{{Grigis}}, \binits{P.}},
\bauthor{\bsnm{{Davey}}, \binits{A.}},
\bauthor{\bsnm{{Martens}}, \binits{P.}},
\bauthor{\bsnm{{Testa}}, \binits{P.}},
\bauthor{\bsnm{{Timmons}}, \binits{R.}},
\bauthor{\bsnm{{Su}}, \binits{Y.}},
\bauthor{\bsnm{{SDO Feature Finding Team}}}:
\byear{2010},
\bctitle{{The SDO flare detective}}.
In: \bbtitle{Bull. Amer. Astron. Soc.}
\bseriesno{41},
\bfpage{402.08}.
\end{bchapter}
\endbibitem

\bibitem[\protect\citeauthoryear{{Harra} \textit{et~al.}}{2011}]{Harra:2011hc}
\begin{barticle}
\bauthor{\bsnm{{Harra}}, \binits{L.K.}},
\bauthor{\bsnm{{Sterling}}, \binits{A.C.}},
\bauthor{\bsnm{{G{\"o}m{\"o}ry}}, \binits{P.}},
\bauthor{\bsnm{{Veronig}}, \binits{A.}}:
\byear{2011},
\batitle{{Spectroscopic Observations of a Coronal Moreton Wave}}.
\bjtitle{\apjl}
\bvolume{737},
\bfpage{L4}.
doi:\doiurl{10.1088/2041-8205/737/1/L4}.
\end{barticle}
\endbibitem

\bibitem[\protect\citeauthoryear{{Kaiser}
  \textit{et~al.}}{2008}]{Kaiser:2008ij}
\begin{barticle}
\bauthor{\bsnm{{Kaiser}}, \binits{M.L.}},
\bauthor{\bsnm{{Kucera}}, \binits{T.A.}},
\bauthor{\bsnm{{Davila}}, \binits{J.M.}},
\bauthor{\bsnm{{St.~Cyr}}, \binits{O.C.}},
\bauthor{\bsnm{{Guhathakurta}}, \binits{M.}},
\bauthor{\bsnm{{Christian}}, \binits{E.}}:
\byear{2008},
\batitle{{The STEREO Mission: An Introduction}}.
\bjtitle{\ssr}
\bvolume{136},
\bfpage{5}\,--\,\blpage{16}.
doi:\doiurl{10.1007/s11214-007-9277-0}.
\end{barticle}
\endbibitem

\bibitem[\protect\citeauthoryear{{Lemen} \textit{et~al.}}{2012}]{Lemen:2012bs}
\begin{barticle}
\bauthor{\bsnm{{Lemen}}, \binits{J.R.}},
\bauthor{\bsnm{{Title}}, \binits{A.M.}},
\bauthor{\bsnm{{Akin}}, \binits{D.J.}},
\bauthor{\bsnm{{Boerner}}, \binits{P.F.}},
\bauthor{\bsnm{{Chou}}, \binits{C.}},
\bauthor{\bsnm{{Drake}}, \binits{J.F.}},
\bauthor{\bsnm{{Duncan}}, \binits{D.W.}},
\bauthor{\bsnm{{Edwards}}, \binits{C.G.}},
\bauthor{\bsnm{{Friedlaender}}, \binits{F.M.}},
\bauthor{\bsnm{{Heyman}}, \binits{G.F.}},
\bauthor{\bsnm{{Hurlburt}}, \binits{N.E.}},
\bauthor{\bsnm{{Katz}}, \binits{N.L.}},
\bauthor{\bsnm{{Kushner}}, \binits{G.D.}},
\bauthor{\bsnm{{Levay}}, \binits{M.}},
\bauthor{\bsnm{{Lindgren}}, \binits{R.W.}},
\bauthor{\bsnm{{Mathur}}, \binits{D.P.}},
\bauthor{\bsnm{{McFeaters}}, \binits{E.L.}},
\bauthor{\bsnm{{Mitchell}}, \binits{S.}},
\bauthor{\bsnm{{Rehse}}, \binits{R.A.}},
\bauthor{\bsnm{{Schrijver}}, \binits{C.J.}},
\bauthor{\bsnm{{Springer}}, \binits{L.A.}},
\bauthor{\bsnm{{Stern}}, \binits{R.A.}},
\bauthor{\bsnm{{Tarbell}}, \binits{T.D.}},
\bauthor{\bsnm{{Wuelser}}, \binits{J.-P.}},
\bauthor{\bsnm{{Wolfson}}, \binits{C.J.}},
\bauthor{\bsnm{{Yanari}}, \binits{C.}},
\bauthor{\bsnm{{Bookbinder}}, \binits{J.A.}},
\bauthor{\bsnm{{Cheimets}}, \binits{P.N.}},
\bauthor{\bsnm{{Caldwell}}, \binits{D.}},
\bauthor{\bsnm{{Deluca}}, \binits{E.E.}},
\bauthor{\bsnm{{Gates}}, \binits{R.}},
\bauthor{\bsnm{{Golub}}, \binits{L.}},
\bauthor{\bsnm{{Park}}, \binits{S.}},
\bauthor{\bsnm{{Podgorski}}, \binits{W.A.}},
\bauthor{\bsnm{{Bush}}, \binits{R.I.}},
\bauthor{\bsnm{{Scherrer}}, \binits{P.H.}},
\bauthor{\bsnm{{Gummin}}, \binits{M.A.}},
\bauthor{\bsnm{{Smith}}, \binits{P.}},
\bauthor{\bsnm{{Auker}}, \binits{G.}},
\bauthor{\bsnm{{Jerram}}, \binits{P.}},
\bauthor{\bsnm{{Pool}}, \binits{P.}},
\bauthor{\bsnm{{Soufli}}, \binits{R.}},
\bauthor{\bsnm{{Windt}}, \binits{D.L.}},
\bauthor{\bsnm{{Beardsley}}, \binits{S.}},
\bauthor{\bsnm{{Clapp}}, \binits{M.}},
\bauthor{\bsnm{{Lang}}, \binits{J.}},
\bauthor{\bsnm{{Waltham}}, \binits{N.}}:
\byear{2012},
\batitle{{The Atmospheric Imaging Assembly (AIA) on the Solar Dynamics
  Observatory (SDO)}}.
\bjtitle{\solphys}
\bvolume{275},
\bfpage{17}\,--\,\blpage{40}.
doi:\doiurl{10.1007/s11207-011-9776-8}.
\end{barticle}
\endbibitem

\bibitem[\protect\citeauthoryear{{Long}, {DeLuca}, and
  {Gallagher}}{2011}]{Long:2011dz}
\begin{barticle}
\bauthor{\bsnm{{Long}}, \binits{D.M.}},
\bauthor{\bsnm{{DeLuca}}, \binits{E.E.}},
\bauthor{\bsnm{{Gallagher}}, \binits{P.T.}}:
\byear{2011},
\batitle{{The Wave Properties of Coronal Bright Fronts Observed Using
  SDO/AIA}}.
\bjtitle{\apjl}
\bvolume{741},
\bfpage{L21}.
doi:\doiurl{10.1088/2041-8205/741/1/L21}.
\end{barticle}
\endbibitem

\bibitem[\protect\citeauthoryear{{Long} \textit{et~al.}}{2011}]{Long:2011fv}
\begin{barticle}
\bauthor{\bsnm{{Long}}, \binits{D.M.}},
\bauthor{\bsnm{{Gallagher}}, \binits{P.T.}},
\bauthor{\bsnm{{McAteer}}, \binits{R.T.J.}},
\bauthor{\bsnm{{Bloomfield}}, \binits{D.S.}}:
\byear{2011},
\batitle{{Deceleration and dispersion of large-scale coronal bright fronts}}.
\bjtitle{\aap}
\bvolume{531},
\bfpage{A42}.
doi:\doiurl{10.1051/0004-6361/201015879}.
\end{barticle}
\endbibitem

\bibitem[\protect\citeauthoryear{{Long} \textit{et~al.}}{2013}]{Long:2013fu}
\begin{botherref}
\oauthor{\bsnm{{Long}}, \binits{D.M.}},
\oauthor{\bsnm{{Williams}}, \binits{D.R.}},
\oauthor{\bsnm{{R{\'e}gnier}}, \binits{S.}},
\oauthor{\bsnm{{Harra}}, \binits{L.K.}}:
2013,
{Measuring the Magnetic-Field Strength of the Quiet Solar Corona Using ''EIT
  Waves''}.
\textit{\solphys}.
doi:\doiurl{10.1007/s11207-013-0331-7}.
\end{botherref}
\endbibitem

\bibitem[\protect\citeauthoryear{{Martens}
  \textit{et~al.}}{2012}]{Martens:2012kl}
\begin{barticle}
\bauthor{\bsnm{{Martens}}, \binits{P.C.H.}},
\bauthor{\bsnm{{Attrill}}, \binits{G.D.R.}},
\bauthor{\bsnm{{Davey}}, \binits{A.R.}},
\bauthor{\bsnm{{Engell}}, \binits{A.}},
\bauthor{\bsnm{{Farid}}, \binits{S.}},
\bauthor{\bsnm{{Grigis}}, \binits{P.C.}},
\bauthor{\bsnm{{Kasper}}, \binits{J.}},
\bauthor{\bsnm{{Korreck}}, \binits{K.}},
\bauthor{\bsnm{{Saar}}, \binits{S.H.}},
\bauthor{\bsnm{{Savcheva}}, \binits{A.}},
\bauthor{\bsnm{{Su}}, \binits{Y.}},
\bauthor{\bsnm{{Testa}}, \binits{P.}},
\bauthor{\bsnm{{Wills-Davey}}, \binits{M.}},
\bauthor{\bsnm{{Bernasconi}}, \binits{P.N.}},
\bauthor{\bsnm{{Raouafi}}, \binits{N.-E.}},
\bauthor{\bsnm{{Delouille}}, \binits{V.A.}},
\bauthor{\bsnm{{Hochedez}}, \binits{J.F.}},
\bauthor{\bsnm{{Cirtain}}, \binits{J.W.}},
\bauthor{\bsnm{{Deforest}}, \binits{C.E.}},
\bauthor{\bsnm{{Angryk}}, \binits{R.A.}},
\bauthor{\bsnm{{de Moortel}}, \binits{I.}},
\bauthor{\bsnm{{Wiegelmann}}, \binits{T.}},
\bauthor{\bsnm{{Georgoulis}}, \binits{M.K.}},
\bauthor{\bsnm{{McAteer}}, \binits{R.T.J.}},
\bauthor{\bsnm{{Timmons}}, \binits{R.P.}}:
\byear{2012},
\batitle{{Computer Vision for the Solar Dynamics Observatory (SDO)}}.
\bjtitle{\solphys}
\bvolume{275},
\bfpage{79}\,--\,\blpage{113}.
doi:\doiurl{10.1007/s11207-010-9697-y}.
\end{barticle}
\endbibitem

\bibitem[\protect\citeauthoryear{{Moses} \textit{et~al.}}{1997}]{Moses:1997qa}
\begin{barticle}
\bauthor{\bsnm{{Moses}}, \binits{D.}},
\bauthor{\bsnm{{Clette}}, \binits{F.}},
\bauthor{\bsnm{{Delaboudini{\`e}re}}, \binits{J.-P.}},
\bauthor{\bsnm{{Artzner}}, \binits{G.E.}},
\bauthor{\bsnm{{Bougnet}}, \binits{M.}},
\bauthor{\bsnm{{Brunaud}}, \binits{J.}},
\bauthor{\bsnm{{Carabetian}}, \binits{C.}},
\bauthor{\bsnm{{Gabriel}}, \binits{A.H.}},
\bauthor{\bsnm{{Hochedez}}, \binits{J.F.}},
\bauthor{\bsnm{{Millier}}, \binits{F.}},
\bauthor{\bsnm{{Song}}, \binits{X.Y.}},
\bauthor{\bsnm{{Au}}, \binits{B.}},
\bauthor{\bsnm{{Dere}}, \binits{K.P.}},
\bauthor{\bsnm{{Howard}}, \binits{R.A.}},
\bauthor{\bsnm{{Kreplin}}, \binits{R.}},
\bauthor{\bsnm{{Michels}}, \binits{D.J.}},
\bauthor{\bsnm{{Defise}}, \binits{J.M.}},
\bauthor{\bsnm{{Jamar}}, \binits{C.}},
\bauthor{\bsnm{{Rochus}}, \binits{P.}},
\bauthor{\bsnm{{Chauvineau}}, \binits{J.P.}},
\bauthor{\bsnm{{Marioge}}, \binits{J.P.}},
\bauthor{\bsnm{{Catura}}, \binits{R.C.}},
\bauthor{\bsnm{{Lemen}}, \binits{J.R.}},
\bauthor{\bsnm{{Shing}}, \binits{L.}},
\bauthor{\bsnm{{Stern}}, \binits{R.A.}},
\bauthor{\bsnm{{Gurman}}, \binits{J.B.}},
\bauthor{\bsnm{{Neupert}}, \binits{W.M.}},
\bauthor{\bsnm{{Newmark}}, \binits{J.}},
\bauthor{\bsnm{{Thompson}}, \binits{B.}},
\bauthor{\bsnm{{Maucherat}}, \binits{A.}},
\bauthor{\bsnm{{Portier-Fozzani}}, \binits{F.}},
\bauthor{\bsnm{{Berghmans}}, \binits{D.}},
\bauthor{\bsnm{{Cugnon}}, \binits{P.}},
\bauthor{\bsnm{{van Dessel}}, \binits{E.L.}},
\bauthor{\bsnm{{Gabryl}}, \binits{J.R.}}:
\byear{1997},
\batitle{{EIT Observations of the Extreme Ultraviolet Sun}}.
\bjtitle{\solphys}
\bvolume{175},
\bfpage{571}\,--\,\blpage{599}.
doi:\doiurl{10.1023/A:1004902913117}.
\end{barticle}
\endbibitem

\bibitem[\protect\citeauthoryear{{Muhr} \textit{et~al.}}{2011}]{Muhr:2011pi}
\begin{barticle}
\bauthor{\bsnm{{Muhr}}, \binits{N.}},
\bauthor{\bsnm{{Veronig}}, \binits{A.M.}},
\bauthor{\bsnm{{Kienreich}}, \binits{I.W.}},
\bauthor{\bsnm{{Temmer}}, \binits{M.}},
\bauthor{\bsnm{{Vr{\v s}nak}}, \binits{B.}}:
\byear{2011},
\batitle{{Analysis of Characteristic Parameters of Large-scale Coronal Waves
  Observed by the Solar-Terrestrial Relations Observatory/Extreme Ultraviolet
  Imager}}.
\bjtitle{\apj}
\bvolume{739},
\bfpage{89}.
doi:\doiurl{10.1088/0004-637X/739/2/89}.
\end{barticle}
\endbibitem

\bibitem[\protect\citeauthoryear{{Nitta} \textit{et~al.}}{2013}]{Nitta:2013kc}
\begin{barticle}
\bauthor{\bsnm{{Nitta}}, \binits{N.V.}},
\bauthor{\bsnm{{Schrijver}}, \binits{C.J.}},
\bauthor{\bsnm{{Title}}, \binits{A.M.}},
\bauthor{\bsnm{{Liu}}, \binits{W.}}:
\byear{2013},
\batitle{{Large-scale Coronal Propagating Fronts in Solar Eruptions as Observed
  by the Atmospheric Imaging Assembly on Board the Solar Dynamics Observatory -
  an Ensemble Study}}.
\bjtitle{\apj}
\bvolume{776},
\bfpage{58}.
doi:\doiurl{10.1088/0004-637X/776/1/58}.
\end{barticle}
\endbibitem

\bibitem[\protect\citeauthoryear{{Olmedo}
  \textit{et~al.}}{2012}]{Olmedo:2012ff}
\begin{barticle}
\bauthor{\bsnm{{Olmedo}}, \binits{O.}},
\bauthor{\bsnm{{Vourlidas}}, \binits{A.}},
\bauthor{\bsnm{{Zhang}}, \binits{J.}},
\bauthor{\bsnm{{Cheng}}, \binits{X.}}:
\byear{2012},
\batitle{{Secondary Waves and/or the ''Reflection'' from and ''Transmission''
  through a Coronal Hole of an Extreme Ultraviolet Wave Associated with the
  2011 February 15 X2.2 Flare Observed with SDO/AIA and STEREO/EUVI}}.
\bjtitle{\apj}
\bvolume{756},
\bfpage{143}.
doi:\doiurl{10.1088/0004-637X/756/2/143}.
\end{barticle}
\endbibitem

\bibitem[\protect\citeauthoryear{{Pesnell}, {Thompson}, and
  {Chamberlin}}{2012}]{Pesnell:2012lh}
\begin{barticle}
\bauthor{\bsnm{{Pesnell}}, \binits{W.D.}},
\bauthor{\bsnm{{Thompson}}, \binits{B.J.}},
\bauthor{\bsnm{{Chamberlin}}, \binits{P.C.}}:
\byear{2012},
\batitle{{The Solar Dynamics Observatory (SDO)}}.
\bjtitle{\solphys}
\bvolume{275},
\bfpage{3}\,--\,\blpage{15}.
doi:\doiurl{10.1007/s11207-011-9841-3}.
\end{barticle}
\endbibitem

\bibitem[\protect\citeauthoryear{{Podladchikova} and
  {Berghmans}}{2005}]{Podladchikova:2005ye}
\begin{barticle}
\bauthor{\bsnm{{Podladchikova}}, \binits{O.}},
\bauthor{\bsnm{{Berghmans}}, \binits{D.}}:
\byear{2005},
\batitle{{Automated Detection Of Eit Waves And Dimmings}}.
\bjtitle{\solphys}
\bvolume{228},
\bfpage{265}\,--\,\blpage{284}.
doi:\doiurl{10.1007/s11207-005-5373-z}.
\end{barticle}
\endbibitem

\bibitem[\protect\citeauthoryear{{Podladchikova}
  \textit{et~al.}}{2010}]{Podladchikova:2010fu}
\begin{barticle}
\bauthor{\bsnm{{Podladchikova}}, \binits{O.}},
\bauthor{\bsnm{{Vourlidas}}, \binits{A.}},
\bauthor{\bsnm{{Van der Linden}}, \binits{R.A.M.}},
\bauthor{\bsnm{{W{\"u}lser}}, \binits{J.-P.}},
\bauthor{\bsnm{{Patsourakos}}, \binits{S.}}:
\byear{2010},
\batitle{{Extreme Ultraviolet Observations and Analysis of Micro-Eruptions and
  Their Associated Coronal Waves}}.
\bjtitle{\apj}
\bvolume{709},
\bfpage{369}\,--\,\blpage{376}.
doi:\doiurl{10.1088/0004-637X/709/1/369}.
\end{barticle}
\endbibitem

\bibitem[\protect\citeauthoryear{{Schrijver}
  \textit{et~al.}}{2011}]{Schrijver:2011qo}
\begin{barticle}
\bauthor{\bsnm{{Schrijver}}, \binits{C.J.}},
\bauthor{\bsnm{{Aulanier}}, \binits{G.}},
\bauthor{\bsnm{{Title}}, \binits{A.M.}},
\bauthor{\bsnm{{Pariat}}, \binits{E.}},
\bauthor{\bsnm{{Delann{\'e}e}}, \binits{C.}}:
\byear{2011},
\batitle{{The 2011 February 15 X2 Flare, Ribbons, Coronal Front, and Mass
  Ejection: Interpreting the Three-dimensional Views from the Solar Dynamics
  Observatory and STEREO Guided by Magnetohydrodynamic Flux-rope Modeling}}.
\bjtitle{\apj}
\bvolume{738},
\bfpage{167}.
doi:\doiurl{10.1088/0004-637X/738/2/167}.
\end{barticle}
\endbibitem

\bibitem[\protect\citeauthoryear{{Shen} \textit{et~al.}}{2013}]{Shen:2013tw}
\begin{barticle}
\bauthor{\bsnm{{Shen}}, \binits{Y.}},
\bauthor{\bsnm{{Liu}}, \binits{Y.}},
\bauthor{\bsnm{{Su}}, \binits{J.}},
\bauthor{\bsnm{{Li}}, \binits{H.}},
\bauthor{\bsnm{{Zhao}}, \binits{R.}},
\bauthor{\bsnm{{Tian}}, \binits{Z.}},
\bauthor{\bsnm{{Ichimoto}}, \binits{K.}},
\bauthor{\bsnm{{Shibata}}, \binits{K.}}:
\byear{2013},
\batitle{{Diffraction, Refraction, and Reflection of an Extreme-ultraviolet
  Wave Observed during Its Interactions with Remote Active Regions}}.
\bjtitle{\apjl}
\bvolume{773},
\bfpage{L33}.
doi:\doiurl{10.1088/2041-8205/773/2/L33}.
\end{barticle}
\endbibitem

\bibitem[\protect\citeauthoryear{{Thompson} and
  {Myers}}{2009}]{Thompson:2009il}
\begin{barticle}
\bauthor{\bsnm{{Thompson}}, \binits{B.J.}},
\bauthor{\bsnm{{Myers}}, \binits{D.C.}}:
\byear{2009},
\batitle{{A Catalog of Coronal ''EIT Wave'' Transients}}.
\bjtitle{\apjs}
\bvolume{183},
\bfpage{225}\,--\,\blpage{243}.
doi:\doiurl{10.1088/0067-0049/183/2/225}.
\end{barticle}
\endbibitem

\bibitem[\protect\citeauthoryear{{Thompson}
  \textit{et~al.}}{1999}]{Thompson:1999zt}
\begin{barticle}
\bauthor{\bsnm{{Thompson}}, \binits{B.J.}},
\bauthor{\bsnm{{Gurman}}, \binits{J.B.}},
\bauthor{\bsnm{{Neupert}}, \binits{W.M.}},
\bauthor{\bsnm{{Newmark}}, \binits{J.S.}},
\bauthor{\bsnm{{Delaboudini{\`e}re}}, \binits{J.-P.}},
\bauthor{\bsnm{{St.~Cyr}}, \binits{O.C.}},
\bauthor{\bsnm{{Stezelberger}}, \binits{S.}},
\bauthor{\bsnm{{Dere}}, \binits{K.P.}},
\bauthor{\bsnm{{Howard}}, \binits{R.A.}},
\bauthor{\bsnm{{Michels}}, \binits{D.J.}}:
\byear{1999},
\batitle{{SOHO/EIT Observations of the 1997 April 7 Coronal Transient: Possible
  Evidence of Coronal Moreton Waves}}.
\bjtitle{\apjl}
\bvolume{517},
\bfpage{L151}\,--\,\blpage{L154}.
doi:\doiurl{10.1086/312030}.
\end{barticle}
\endbibitem

\bibitem[\protect\citeauthoryear{{Uchida}}{1968}]{Uchida:1968gb}
\begin{barticle}
\bauthor{\bsnm{{Uchida}}, \binits{Y.}}:
\byear{1968},
\batitle{{Propagation of Hydromagnetic Disturbances in the Solar Corona and
  Moreton's Wave Phenomenon}}.
\bjtitle{\solphys}
\bvolume{4},
\bfpage{30}\,--\,\blpage{44}.
doi:\doiurl{10.1007/BF00146996}.
\end{barticle}
\endbibitem

\bibitem[\protect\citeauthoryear{{Uchida}}{1970}]{Uchida:1970jl}
\begin{barticle}
\bauthor{\bsnm{{Uchida}}, \binits{Y.}}:
\byear{1970},
\batitle{{Diagnosis of Coronal Magnetic Structure by Flare-Associated
  Hydromagnetic Disturbances}}.
\bjtitle{\pasj}
\bvolume{22},
\bfpage{341}.
\end{barticle}
\endbibitem

\bibitem[\protect\citeauthoryear{{Veronig}
  \textit{et~al.}}{2006}]{Veronig:2006cq}
\begin{barticle}
\bauthor{\bsnm{{Veronig}}, \binits{A.M.}},
\bauthor{\bsnm{{Temmer}}, \binits{M.}},
\bauthor{\bsnm{{Vr{\v s}nak}}, \binits{B.}},
\bauthor{\bsnm{{Thalmann}}, \binits{J.K.}}:
\byear{2006},
\batitle{{Interaction of a Moreton/EIT Wave and a Coronal Hole}}.
\bjtitle{\apj}
\bvolume{647},
\bfpage{1466}\,--\,\blpage{1471}.
doi:\doiurl{10.1086/505456}.
\end{barticle}
\endbibitem

\bibitem[\protect\citeauthoryear{{Veronig}
  \textit{et~al.}}{2011}]{Veronig:2011mb}
\begin{barticle}
\bauthor{\bsnm{{Veronig}}, \binits{A.M.}},
\bauthor{\bsnm{{Gomory}}, \binits{P.}},
\bauthor{\bsnm{{Kienreich}}, \binits{I.W.}},
\bauthor{\bsnm{{Muhr}}, \binits{N.}},
\bauthor{\bsnm{{Vrsnak}}, \binits{B.}},
\bauthor{\bsnm{{Temmer}}, \binits{M.}},
\bauthor{\bsnm{{Warren}}, \binits{H.P.}}:
\byear{2011},
\batitle{{Plasma diagnostics of an EIT wave observed by Hinode/EIS and
  SDO/AIA.}}
\bjtitle{\apjl}
\bvolume{743},
\bfpage{L10}.
doi:\doiurl{10.1088/2041-8205/743/1/L10}.
\end{barticle}
\endbibitem

\bibitem[\protect\citeauthoryear{{Warmuth} and {Mann}}{2011}]{Warmuth:2011kh}
\begin{barticle}
\bauthor{\bsnm{{Warmuth}}, \binits{A.}},
\bauthor{\bsnm{{Mann}}, \binits{G.}}:
\byear{2011},
\batitle{{Kinematical evidence for physically different classes of large-scale
  coronal EUV waves}}.
\bjtitle{\aap}
\bvolume{532},
\bfpage{A151}.
doi:\doiurl{10.1051/0004-6361/201116685}.
\end{barticle}
\endbibitem

\bibitem[\protect\citeauthoryear{{West} \textit{et~al.}}{2011}]{West:2011rq}
\begin{barticle}
\bauthor{\bsnm{{West}}, \binits{M.J.}},
\bauthor{\bsnm{{Zhukov}}, \binits{A.N.}},
\bauthor{\bsnm{{Dolla}}, \binits{L.}},
\bauthor{\bsnm{{Rodriguez}}, \binits{L.}}:
\byear{2011},
\batitle{{Coronal Seismology Using EIT Waves: Estimation of the Coronal
  Magnetic Field Strength in the Quiet Sun}}.
\bjtitle{\apj}
\bvolume{730},
\bfpage{122}.
doi:\doiurl{10.1088/0004-637X/730/2/122}.
\end{barticle}
\endbibitem

\bibitem[\protect\citeauthoryear{{Wills-Davey}}{2006}]{Wills-Davey:2006ss}
\begin{barticle}
\bauthor{\bsnm{{Wills-Davey}}, \binits{M.J.}}:
\byear{2006},
\batitle{{Tracking Large-Scale Propagating Coronal Wave Fronts (EIT Waves)
  using Automated Methods}}.
\bjtitle{\apj}
\bvolume{645},
\bfpage{757}\,--\,\blpage{765}.
doi:\doiurl{10.1086/504144}.
\end{barticle}
\endbibitem

\bibitem[\protect\citeauthoryear{{Wills-Davey} and
  {Attrill}}{2009}]{Wills-Davey:2009qc}
\begin{barticle}
\bauthor{\bsnm{{Wills-Davey}}, \binits{M.J.}},
\bauthor{\bsnm{{Attrill}}, \binits{G.D.R.}}:
\byear{2009},
\batitle{{EIT Waves: A Changing Understanding over a Solar Cycle}}.
\bjtitle{\ssr}
\bvolume{149},
\bfpage{325}\,--\,\blpage{353}.
doi:\doiurl{10.1007/s11214-009-9612-8}.
\end{barticle}
\endbibitem

\bibitem[\protect\citeauthoryear{{Wills-Davey} and
  {Thompson}}{1999}]{Wills-Davey:1999fc}
\begin{barticle}
\bauthor{\bsnm{{Wills-Davey}}, \binits{M.J.}},
\bauthor{\bsnm{{Thompson}}, \binits{B.J.}}:
\byear{1999},
\batitle{{Observations of a Propagating Disturbance in TRACE}}.
\bjtitle{\solphys}
\bvolume{190},
\bfpage{467}\,--\,\blpage{483}.
doi:\doiurl{10.1023/A:1005201500675}.
\end{barticle}
\endbibitem

\bibitem[\protect\citeauthoryear{{Wills-Davey}, {DeForest}, and
  {Stenflo}}{2007}]{Wills-Davey:2007mw}
\begin{barticle}
\bauthor{\bsnm{{Wills-Davey}}, \binits{M.J.}},
\bauthor{\bsnm{{DeForest}}, \binits{C.E.}},
\bauthor{\bsnm{{Stenflo}}, \binits{J.O.}}:
\byear{2007},
\batitle{{Are ``EIT Waves'' Fast-Mode MHD Waves?}}
\bjtitle{\apj}
\bvolume{664},
\bfpage{556}\,--\,\blpage{562}.
doi:\doiurl{10.1086/519013}.
\end{barticle}
\endbibitem

\bibitem[\protect\citeauthoryear{{Wuelser}
  \textit{et~al.}}{2004}]{Wuelser:2004bd}
\begin{bchapter}
\bauthor{\bsnm{{Wuelser}}, \binits{J.-P.}},
\bauthor{\bsnm{{Lemen}}, \binits{J.R.}},
\bauthor{\bsnm{{Tarbell}}, \binits{T.D.}},
\bauthor{\bsnm{{Wolfson}}, \binits{C.J.}},
\bauthor{\bsnm{{Cannon}}, \binits{J.C.}},
\bauthor{\bsnm{{Carpenter}}, \binits{B.A.}},
\bauthor{\bsnm{{Duncan}}, \binits{D.W.}},
\bauthor{\bsnm{{Gradwohl}}, \binits{G.S.}},
\bauthor{\bsnm{{Meyer}}, \binits{S.B.}},
\bauthor{\bsnm{{Moore}}, \binits{A.S.}},
\bauthor{\bsnm{{Navarro}}, \binits{R.L.}},
\bauthor{\bsnm{{Pearson}}, \binits{J.D.}},
\bauthor{\bsnm{{Rossi}}, \binits{G.R.}},
\bauthor{\bsnm{{Springer}}, \binits{L.A.}},
\bauthor{\bsnm{{Howard}}, \binits{R.A.}},
\bauthor{\bsnm{{Moses}}, \binits{J.D.}},
\bauthor{\bsnm{{Newmark}}, \binits{J.S.}},
\bauthor{\bsnm{{Delaboudiniere}}, \binits{J.-P.}},
\bauthor{\bsnm{{Artzner}}, \binits{G.E.}},
\bauthor{\bsnm{{Auchere}}, \binits{F.}},
\bauthor{\bsnm{{Bougnet}}, \binits{M.}},
\bauthor{\bsnm{{Bouyries}}, \binits{P.}},
\bauthor{\bsnm{{Bridou}}, \binits{F.}},
\bauthor{\bsnm{{Clotaire}}, \binits{J.-Y.}},
\bauthor{\bsnm{{Colas}}, \binits{G.}},
\bauthor{\bsnm{{Delmotte}}, \binits{F.}},
\bauthor{\bsnm{{Jerome}}, \binits{A.}},
\bauthor{\bsnm{{Lamare}}, \binits{M.}},
\bauthor{\bsnm{{Mercier}}, \binits{R.}},
\bauthor{\bsnm{{Mullot}}, \binits{M.}},
\bauthor{\bsnm{{Ravet}}, \binits{M.-F.}},
\bauthor{\bsnm{{Song}}, \binits{X.}},
\bauthor{\bsnm{{Bothmer}}, \binits{V.}},
\bauthor{\bsnm{{Deutsch}}, \binits{W.}}:
\byear{2004},
\bctitle{{EUVI: the STEREO-SECCHI extreme ultraviolet imager}}.
In: \beditor{\bsnm{{Fineschi}}, \binits{S.}},
\beditor{\bsnm{{Gummin}}, \binits{M.A.}} (eds.)
\bbtitle{SPIE Conf. Ser.}
\bseriesno{5171},
\bfpage{111}\,--\,\blpage{122}.
doi:\doiurl{10.1117/12.506877}.
\end{bchapter}
\endbibitem

\bibitem[\protect\citeauthoryear{{Zhukov}}{2011}]{Zhukov:2011ud}
\begin{barticle}
\bauthor{\bsnm{{Zhukov}}, \binits{A.N.}}:
\byear{2011},
\batitle{{EIT wave observations and modeling in the STEREO era}}.
\bjtitle{Journal of Atmospheric and Solar-Terrestrial Physics}
\bvolume{73},
\bfpage{1096}\,--\,\blpage{1116}.
doi:\doiurl{10.1016/j.jastp.2010.11.030}.
\end{barticle}
\endbibitem

\bibitem[\protect\citeauthoryear{{Zhukov} and
  {Auch{\`e}re}}{2004}]{Zhukov:2004if}
\begin{barticle}
\bauthor{\bsnm{{Zhukov}}, \binits{A.N.}},
\bauthor{\bsnm{{Auch{\`e}re}}, \binits{F.}}:
\byear{2004},
\batitle{{On the nature of EIT waves, EUV dimmings and their link to CMEs}}.
\bjtitle{\aap}
\bvolume{427},
\bfpage{705}\,--\,\blpage{716}.
doi:\doiurl{10.1051/0004-6361:20040351}.
\end{barticle}
\endbibitem

\end{thebibliography}

\end{article} 
\end{document}